\documentclass[aps,12pt, a4paper]{revtex4}

\usepackage[utf8]{inputenc}
\usepackage[T1]{fontenc}
\usepackage{amsmath}
\usepackage{amsfonts}
\usepackage{amssymb}
\usepackage{graphics,graphicx}

\usepackage{graphicx,color}
\usepackage{subfigure} 
\usepackage{wrapfig} 
\usepackage{epstopdf}
\graphicspath{{figuras/}}

\usepackage[breaklinks=true]{hyperref}
\usepackage{setspace}

\newcommand{\bes}{\begin{subequations}}
\newcommand{\ees}{\end{subequations}}
\def\ben{\begin{eqnarray}}
\def\een{\end{eqnarray}}
\def\be{\begin{equation}}
\def\ee{\end{equation}}

\begin{document}

\title{Complexity of four-dimensional hairy Anti-de-Sitter black holes with a rotating string and shear viscosity in generalized scalar-tensor theories}

\author{Mois\'es Bravo-Gaete}
\email{mbravo-at-ucm.cl} \affiliation{Facultad de Ciencias
B\'asicas, Universidad Cat\'olica del Maule, Casilla 617, Talca,
Chile.}

\author{F. F. Santos}
\email{fabiano.ffs23-at-gmail.com} \affiliation{Departamento de F\'\i sica, Universidade Federal da Para\'iba, Caixa Postal 5008, 58051-970, Jo\~ ao Pessoa, Para\' iba, Brazil\\
Instituto Federal de Educação, Ciência e Tecnologia do Sertão Pernambucano, 56316-686, Campus Petrolina Zona Rural, Pernambuco, Brazil}

\begin{abstract}
In four dimensions, we consider a generalized scalar-tensor theory where the coupling functions only depend on the kinetic term of the scalar field. For this model, we obtain a set of hairy Anti-de-Sitter black hole solutions, allowing us to calculate the computational complexity, according to the Complexity equals Action conjecture. To perform this, the system contains a particle moving on the boundary, corresponding to the insertion of a fundamental string in the bulk. The effect string is given by the Nambu-Goto term, analyzing the time development of this system. Together with the above, we calculate the shear viscosity, where the viscosity/\textcolor{black}{entropy density} ratio can violate the Kovtun-Son-Starinets bound for a suitable choice of coupling functions.
\end{abstract}

\maketitle
\newpage

\section{Introduction}
\label{intro}

In recent years, there is a deep interest in the application to black hole physics from computational complexity  \footnote{
Defined as the number of necessary gates which operate
to produce the target state from the reference  state.} (see for example \cite{Nagasaki:2017kqe,Nagasaki:2018csh,Santos:2020xox,Carmi:2017jqz,Roberts:2016hpo,Stanford:2014jda,Susskind:2014rva}), thanks to the seminal ideas from the quantum information process  \cite{Watrous2008} as well as arguments for the existence of a second law of complexity, behaving similar to the entropy \cite{Brown:2017jil}. In particular, and based on the works  \cite{Almheiri:2012rt,Harlow:2013tf,Susskind:2015toa}, the growth in complexity is related to problems of black holes, such as information problem, the existence of firewalls, or transparency of horizons.

To connect quantum complexity and black holes, in \cite{Brown:2015bva,Brown:2015lvg} the authors provide the Complexity equals Action (CA) conjecture, where the complexity is dual to the bulk action on a certain region of the spacetime denominated as Wheeler-de Witt patch (WdW). A generalization of this conjecture can be performed for systems other than stationary states, such as a system powered by a drag force \cite{Nagasaki:2017kqe}. For this work, the idea is to consider a Wilson line operator located on a five-dimensional Anti-de-Sitter ($\mbox{AdS}_{5}$) spacetime, inserting a fundamental string, being reflected by the addition of a Nambu-Goto (NG) term to the Einstein-Hilbert action together with a cosmological constant. 

This procedure has been explored for the three dimensional case \cite{Santos:2020xox}, studying the effects of the probe string on the Ba\~{n}ados-Teitelboim-Zanelli black hole \cite{Banados:1992wn} given by a special case of the Horndeski theory \cite{Horndeski:1974wa}, given by a Lagrangian
\begin{equation}\label{eq:horntrunc}
 {\mathcal{L}}=\kappa R -2 \Lambda -\gamma X + \beta G_{\mu \nu} \phi^{\mu} \phi^{\nu},   
\end{equation}
where $\kappa$, $\gamma$ and $\beta$ are coupling constant,
$X:=\partial_{\mu}\phi\,\partial^{\mu}\phi$, $\phi_{\mu}:=\nabla_{\mu}\phi$, while that $R$, $G_{\mu \nu}$ and $\Lambda$ are  the scalar curvature, the Einstein tensor and the cosmological constant respectively. For the theory (\ref{eq:horntrunc}), the inclusion of the fundamental string is given through the NG term. Here, although the string is stationary,  provides maximum growth in complexity where the action growth is influenced by the parameters of the theory. 

On the other hand, AdS planar black holes allow us to analyze the effects on shear viscosity, thanks to the gravity/gauge duality \cite{Maldacena:1997re,Gubser:1998bc,Witten:1998qj} and by effective coupling constants of the transverse graviton on the location of the event horizon \cite{Iqbal:2008by},  where the fluid is characterized by the production of entropy \cite{Brito:2019ose}. For some common substances (for example helium, nitrogen, and water) the viscosity/ \textcolor{black}{entropy density} $(\eta/s)$ ratio  is always substantially greater than its value in dual gravity theories, which reads as \cite{Kovtun:2003wp,Kovtun:2004de,Son:2002sd}
\begin{equation}\label{eq:kss}
\frac{\eta}{s}\geq \frac{1}{4\pi},   
\end{equation}
known as the Kovtun-Son-Starinets (KSS) bound. This value has been shown in a variety of cases (see \cite{Buchel:2003tz,Buchel:2004qq,Benincasa:2006fu,Landsteiner:2007bd}). However, there are concrete examples where the KSS bound (\ref{eq:kss}) is violated, for example, via 
gravitational theories such as the five-dimensional Einstein-Hilbert Gauss-Bonnet model \cite{Kats:2007mq,Brigante:2007nu}  as well as scalar-tensor theories \cite{Brito:2019ose}, \cite{Feng:2015oea}, where the Lagrangian is given by (\ref{eq:horntrunc}).

The above leads us to the motivations of the present paper, which is to explore hairy black holes solutions in four dimensions and their thermodynamics, allowing us to obtain the CA conjecture following the procedure performed in \cite{Nagasaki:2017kqe,Nagasaki:2018csh}, considering a Wilson line operator located in the AdS$_{4}$ spacetime, by inserting a fundamental string, described by the addition of the NG term and computing the growth of the complexity of a dissipative system. Together with the above, constructing a Noether charge with a suitable election of a space-like Killing vector, we can  compute the $\eta/s$ ratio  by using the infrared data on the black hole event horizon \cite{Fan:2018qnt}, greatly simplifying the steps unlike the traditional procedures \cite{Son:2002sd,Brigante:2007nu}. For all these calculations, the action is described by the four-dimensional  Degenerate-Higher-Order-Scalar-Tensor (DHOST) theory
\begin{subequations}
\begin{eqnarray}
S[g_{\mu \nu},\phi]&=&\int d^{4}x\sqrt{-g}\mathcal{L},\label{action} \\
\mathcal{L}&=&Z(X)+G(X)R+\sum_{i=2}^{5} A_{i} {\mathcal{L}}_{i},\label{lagrangian}
\end{eqnarray}
\end{subequations}
where
\begin{eqnarray}
{\mathcal{L}}_{2}&:=&\left((\Box\phi)^2-\phi_{\mu\nu}\phi^{\mu\nu}\right),\qquad  {\mathcal{L}}_{3}:=\Box\phi\,\phi^{\mu}\phi_{\mu\nu}\phi^{\nu},\\
{\mathcal{L}}_{4}&:=&\phi^{\mu}\phi_{\mu\nu}\phi^{\nu\rho}\phi_{\rho}
,\qquad {\mathcal{L}}_{5}:=\left(\phi^{\mu}\phi_{\mu\nu}\phi^{\nu}\right)^2,\label{Li}
\end{eqnarray}
and the equations of motions with respect to the metric $g_{\mu \nu}$ and the scalar field $\phi$  can be written in the following form
\begin{eqnarray}
\mathcal{E}_{\mu\nu}&:=&{\cal{G}}^{Z}_{\mu\nu}
+{\cal{G}}^{G}_{\mu\nu}+\sum_{i=2}^{5}{\cal{G}}^{(i)}_{\mu\nu}=0,\label{eq:emotion1}\\
\varepsilon_{(\phi)}&=&\nabla_{\mu}
\mathcal{J}^{\mu}=\nabla_{\mu}\left[\frac{\delta \cal{L}}{\delta
(\phi_{\mu})}-\nabla_{\nu}\left(\frac{\delta \cal{L}}{\delta (
\phi_{\mu \nu})}\right)\right]=0,\label{eq:emotion2}
\end{eqnarray}
where  ${\cal{G}}^{Z}_{\mu\nu}
,{\cal{G}}^{G}_{\mu\nu}$, the ${\cal{G}}^{(i)}_{\mu\nu}$'s and $J^{\mu}$ are reported in the Appendix. Here we defined 
$\phi_{\mu\nu}:=\nabla_{\mu}\nabla_{\nu}\phi$.

Recently, the DHOST theory (\ref{action})-(\ref{lagrangian}) has attracted a lot of attention for many reasons, one of them is the simple way to add new degrees of freedom through the introduction of a scalar field $\phi$, being an extension of the Horndeski theory \cite{Horndeski:1974wa},  and avoiding Ostrogradsky instability due to its degeneracy property (see for example \cite{BenAchour:2016fzp,Motohashi:2016ftl}),  \textcolor{black}{as well as allows constructing regular black hole solutions via a Kerr-Schild transformation \cite{Babichev:2020qpr,Baake:2021jzv},  rotating stealth black holes with a cohomogeneity - 1 metric \cite{Baake:2021kyg} and three-dimensional spinning configurations \cite{Baake:2020tgk} }. Here is important to note that the functions $Z, G$ and $A_i$ for $i=\{2,..., 5\}$, are arbitrary
functions of the kinetic term $X$, where for example for $Z(X)=-2 \Lambda$, $G(X)=\kappa$ and $A_{i}=0 \quad \forall i$, the Einstein-Hilbert action together with a cosmological constant $\Lambda$ is recovered, while that for
\begin{equation}\label{eq:sf-s-theory}
A_{i}=0, \mbox{ with } i=\{3,4,5\} \mbox{ and } A_{2}:=-2G_{X},
\end{equation}
where $G_{X}:=dG/dX$,  we have a subclass of the Horndeski theory \cite{Horndeski:1974wa} which enjoys shift symmetry $\phi\to \phi+\mbox{constant}$ and 
$\phi\to -\phi$, being deeply analyzed in \cite{Kobayashi:2014eva}. Just for completeness, the linear case (\ref{eq:horntrunc}) is recovered with (\ref{eq:sf-s-theory}) and 
\begin{eqnarray}
Z(X)&=&-2 \Lambda -\gamma X,\qquad G(X)=\kappa-\left(\frac{\beta}{2}\right) X. \label{eq:special-hordeski}
\end{eqnarray}

The plan of the paper is organized as follows: In Section \ref{v2} we present a  black hole solution considering the theory (\ref{action})-(\ref{lagrangian}), analyzing the thermodynamic quantities in Section \ref{sec.termo}. In  Section \ref{v3}, we calculate the NG term of the string moving on the black hole. In Section \ref{viscosity}, the shear viscosity is computed, obtaining the $\eta/s$ ratio and showing its  dependence on the coupling functions, where the KSS bound can be violated. Finally, Section \ref{v4} is devoted to our conclusions and discussions.

\section{A black hole solution in generalized scalar-tensor theories}\label{v2}

For the generalized scalar-tensor model (\ref{action})-(\ref{lagrangian}), we consider the following four-dimensional metric Ansatz:
\begin{eqnarray}
ds^{2}&=&-h(r)dt^{2}+\frac{dr^{2}}{f(r)}
+r^{2}d\Omega^{2}_{2,\epsilon},\label{7} \\
\phi(t,r)&=&\psi(r)+\phi_0t,\label{phi}
\end{eqnarray}
where
\begin{eqnarray}
d\Omega^{2}_{2,\epsilon} = \left\{ \begin{array}{ll}
dx_{1}^{2}+\sin^{2}(x_{1}) dx_{2}^{2}, & \mbox{for $ \epsilon=1,$}\\
dx_{2}^{2}+\sinh^{2}(x_{1}) dx_{2}^{2}, & \mbox{for $ \epsilon=-1,$}\\
dx_{1}^{2}+dx_{2}^{2}, & \mbox{for $ \epsilon=0,$}
\end{array} \right. 
\end{eqnarray}
and $\phi_0$ is an integration constant. In order to simplify our computations, \textcolor{black}{ given the structure of the metric (\ref{7}) and the scalar field $\phi$ from (\ref{phi}), we suppose that the kinetic term 
\begin{equation}
X=\partial_{\mu}\phi \partial^{\mu}\phi =g^{tt} (\partial_{t}\phi)^2 + g^{rr} (\partial_{r}\phi)^2 =-\frac{\phi_0^2}{h}+f (\psi ')^2,
\end{equation}
is a constant}, this is due to the complexity to compute the NG term, which will be studied in Section \ref{v3}. The above implies that the square of the derivative of the radial component of the scalar field $\phi$ can be cast as
\begin{equation}
(\psi')^{2}=\frac{\phi_0^2}{h f}+\frac{X}{f},
\end{equation}
where $(')$ denotes the derivative respects to the radial coordinate $r$.

In the following lines, to simplify computations and to show in a simple way the solutions, we fix the function $A_5$ as \cite{Baake:2020tgk}
\begin{subequations}
\begin{eqnarray}
A_5=\frac{\left(2A_2+XA_3+4G_X\right)^2}{2X(G+XA_2)}-\frac{A_3+A_4}{X}.
\label{A5}
\end{eqnarray}
Here we note that this relation is not new. In fact, in the literature, we can find gravity theories given by (\ref{action}), which enjoy relations between the coupling functions (see for example \cite{Babichev:2017guv,Chagoya:2018lmv,Charmousis:2019vnf,Baake:2021jzv,Baake:2021kyg}). In what follows, we will consider the Einstein equation (\ref{eq:emotion1}) and defining the functions
\label{defZ}
\begin{eqnarray}
\mathcal{Z}_1 &=& G + X A_2, \label{defZ1}\\
\mathcal{Z}_2 &=& 2 A_2 + X A_3 + 4 G_X. \label{defZ2}
\end{eqnarray}
\end{subequations}
Here, the $(r,r)$-component of the Einstein equations (\ref{eq:emotion1}) together with $\mathcal{J}^{r}=0$ from (\ref{eq:emotion2}), allow to obtain a relation between the metric functions $f$ and $h$ given by
\begin{eqnarray}\label{eq:f}
f={\frac {X h \left( {r}^{2} Z +2\,\epsilon\,G  \right) }{2 [X \mathcal{Z}_{1} (hr)'-\phi_0^{2}
 \left( G -\mathcal{Z}_{1}\right) ]}},
\end{eqnarray}
and the $(t,t)$-component of the Einstein equations yields a second order differential equation for the metric function $h$ which reads
\begin{eqnarray}
&&\{\mathcal{Z}_1[r X \mathcal{Z}_1 (r^2 Z +
2\epsilon G)h''+4\epsilon X \mathcal{Z}_1 G
h'-2 Zr X \mathcal{Z}_1h\nonumber\\
&&-2 Z r \phi_0^{2}(\mathcal{Z}_1 -G)]Xh^2\}\Big{/}\{2 r[X \mathcal{Z}_{1} (hr)'-\phi_0^{2}
 \left( G -\mathcal{Z}_{1}\right)]^2\}=0.
\end{eqnarray}
If we suppose $X \neq 0$ and $\mathcal{Z}_{1}\neq0$, the metric function $h$ takes the form
\begin{eqnarray}\label{eq:h}
h(r)=\left( 6\,G\epsilon+{r}^{2}Z \right) C_{0}-{\frac {M}{r}}+
{\frac {\phi_0^{2} \left( G-\mathcal{Z}_1 \right) }{X \mathcal{Z}_{1}}},
\end{eqnarray}
with $C_0$ and $M$ integrations constants, while that $\mathcal{J}^{t}=0$ from (\ref{eq:emotion2}) is  trivially satisfied. Then, replacing the metric function $h$ from (\ref{eq:h}) in (\ref{eq:f}), we have that
\begin{equation}
\frac{h}{f}=6 \mathcal{Z}_1 C_0=\mbox{constant},
\end{equation}
imposing without loss of generality that
\begin{equation}
C_0=\frac{1}{6 \mathcal{Z}_1}.
\end{equation}
Finally the $(x_i,x_i)$- component of the Einstein equations yields the expression
\begin{eqnarray}\label{eq:mastereq}
r^{2} \mathcal{G}(X)+(\epsilon X+ \phi_0^2)=0,
\end{eqnarray}
where
\begin{equation}\label{eq:mastereq2}
\mathcal{G}(X)=\frac{X \left[4 (\mathcal{Z}_1 Z)_{X}-3 Z \mathcal{Z}_{2}\right]}{4\left[2 (\mathcal{Z}_1 G)_{X}-G \mathcal{Z}_{2}\right]},
\end{equation}
and the solution is given when
\begin{eqnarray}
h(r)=f(r)&=&\epsilon+\frac{Z r^{2}}{6 \, \mathcal{Z}_1}-\frac{M}{r},\label{10}\\
(\psi')^{2}&=&\frac{X (f-\epsilon)}{f^{2}},\qquad \mathcal{G}(X)=0=\epsilon X+ \phi_0^2,\label{11}
\end{eqnarray}
where we suppose that the integration constant $M$, related to the mass black hole $\mathcal{M}$, is always positive. Many commentaries can be made respect to the solution (\ref{eq:mastereq})-(\ref{11}). First that all, the equation (\ref{eq:mastereq}) shows that for the planar base manifold for the event horizon ($\epsilon=0$) only a radial dependence for the scalar field is needed (this is $\phi_0=0$). Together with the above,  for the other cases, the sign of the kinetic term must be adjusted to have $\phi_0^2=-\epsilon X>0$. In addition, this solution resembles the four-dimensional metric with the presence of an effective cosmological constant and a non-trivial expression for the scalar field $\phi(t,r)$. These configurations have been explored in \cite{Babichev:2013cya,Anabalon:2013oea,Bravo-Gaete:2014haa} with the special truncation of the Horndeski theory (\ref{eq:special-hordeski}), in \cite{Kobayashi:2014eva} with the conditions (\ref{eq:sf-s-theory}) and in \cite{Baake:2020tgk} for the three-dimensional situation with the complete action (\ref{action})-(\ref{lagrangian}). Nevertheless, these studies begin with a kinetic term $X=X(r)$, in our situation, {\em{a priori}} we are supposing that this one is constant.  

Given that to analyze the NG-term and the KSS bound the mass as well as the entropy are required quantities, in the following section we will derive the thermodynamic properties.

\section{Thermodynamics of the solution}
\label{sec.termo}
Given the steps performed in the previous section, results interesting to explore the thermodynamics of the solution (\ref{eq:mastereq})-(\ref{11}). As a first computation, we will consider the Hawking Temperature $T$ which reads
\begin{eqnarray}
T=\frac{\kappa}{2 \pi}\Big{|}_{r=r_h}&=&\frac{1}{4 \pi}
\left(\frac{3 r_h}{L^{2}}+\frac{\epsilon}{r_h}\right),\label{1t1}
\end{eqnarray}
constructed by the surface gravity given by
\begin{equation}
\kappa=\sqrt{-\frac{1}{2}\left(\nabla_{\mu} \xi_{\nu}\right)\left(\nabla^{\mu} \xi^{\nu}\right)},
\end{equation}
with a time-like Killing vector {$\partial_{t}=\xi^{\mu}\partial_{\mu}$. Here we have defined
\begin{equation}\label{eq:L2}
L^2=\frac{6 \mathcal{Z}_1}{Z},
\end{equation}
and $r_h$ is the location of the event horizon.

On the other hand, the mass of these configurations will be calculated via the quasilocal formulation described in \cite{Kim:2013zha,Gim:2014nba}, corresponding to an off-shell prescription of the Abbott-Deser-Tekin (ADT) \cite{Abbott:1981ff,Deser:2002rt,Deser:2002jk} procedure. This technique has shown to be useful to compute the mass of black hole solutions with non-standard asymptotically behavior \cite{Herrera-Aguilar:2020iti} and even with quadratic corrections (see for example \cite{Bravo-Gaete:2020ftn,Ayon-Beato:2019kmz,BravoGaete:2017dso,Bravo-Gaete:2021kgt}). The principal elements are the surface term $J^{\mu}$ and the Noether potential $Q^{\mu\nu}$, given by \cite{Baake:2020tgk}
\begin{eqnarray}
J^{\mu}&=&\sqrt{-g}\Big[2\left(P^{\mu (\alpha\beta)
\gamma}\nabla_{\gamma}\delta g_{\alpha\beta}-\delta g_{\alpha\beta} \nabla_{\gamma}P^{\mu(\alpha\beta)\gamma}\right) \nonumber \\
&+&\frac{\delta \cal{L}}{\delta (\phi_{\mu})} \delta
\phi-\nabla_{\nu}\left(\frac{\delta \cal{L}}{\delta (\phi_{\mu
\nu})}\right) \delta \phi
+\frac{\delta \cal{L}}{\delta (\phi_{\mu \nu})} \delta (\phi_{\nu})\nonumber \\
&-&\frac{1}{2}\frac{\delta \cal{L}}{\delta (\phi_{\mu \rho})}
\phi^{\sigma} \,
\delta g_{\sigma \rho}-\frac{1}{2}\frac{\delta \cal{L}}{\delta (\phi_{\rho \mu})}\phi^{\sigma} \,\delta g_{\sigma \rho} \nonumber \\
&+&\frac{1}{2}\frac{\delta \cal{L}}{\delta ( \phi_{\sigma
\rho})}\phi^{\mu}\,\delta g_{\sigma \rho}\Big],\label{eq:surface}\\
Q^{\mu\nu}&=&\sqrt{-g}\,\Big[2P^{\mu\nu\rho\sigma}\nabla_\rho \xi_\sigma -4\xi_\sigma
\nabla_\rho P^{\mu\nu\rho\sigma}+\frac{\delta \cal{L}}{ \delta \phi_{\mu \sigma}} \phi^{\nu} \xi_{\sigma}\nonumber \\
&-&\frac{\delta \cal{L}}{ \delta \phi_{\nu \sigma}} \phi^{\mu}
\xi_{\sigma}\Big], \label{eq:noether}
\end{eqnarray}
where $P^{\mu\nu\lambda\rho}=\delta {\cal{L}}/ \delta R_{\mu\nu\lambda\rho}$, and ${\cal{L}}$ is the Lagrangian given previously in (\ref{lagrangian}). Finally, by using a parameter $\zeta \in [0,1]$, we interpolate between the solution at $\zeta=1$ and the asymptotic one at $\zeta=0$, obtaining \textcolor{black}{the quasilocal charge
\begin{eqnarray}
\mathcal{M}(\xi)=\int_B d^{2}x_{\mu\nu}\left(\Delta Q^{\mu \nu}(\xi)-2\xi^{[\mu} \int^{1}_0 d\zeta J^{\nu]}\right),
\end{eqnarray}
with $\Delta Q^{\mu\nu}(\xi):= Q^{\mu\nu}_{\zeta=1}(\xi)-Q^{\mu\nu}_{\zeta=0}(\xi)$. Via a time-like Killing vector $\xi=\partial_{t}$, we have 
\begin{equation}
\Delta Q^{\mu\nu}(\xi)=(-3G+4\mathcal{Z}_1)\Omega_{2,\epsilon}  ,\qquad  \int^{1}_0 d\zeta J^r= (3G-2\mathcal{Z}_1) M \Omega_{2,\epsilon},
\end{equation}
and the mass $\mathcal{M}$ acquires the structure
\begin{eqnarray}
\label{mass-d4}
\mathcal{M}&=&2 \mathcal{Z}_1 M \Omega_{2,\epsilon}=2 \mathcal{Z}_1 \Omega_{2,\epsilon} \left({\frac {r_h^{3}}{L^{2}}}+\epsilon r_{h}\right),
\end{eqnarray}
where $\Omega_{2,\epsilon}$ is the finite volume of the 2-dimensional compact angular base manifold.}

To compute the entropy, we will consider the Wald procedure
\cite{Wald:1993nt,Iyer:1994ys}. The steps start with the surface term given previously in (\ref{eq:surface}), defining a $1$-form $J_{(1)}=J_{\mu} dx^{\mu}$ and its corresponding Hodge dual ${\Theta}_{(3)}=(-1)* J_{(1)}$. Then, with the equations of motions, we have that
\begin{equation}
J_{(3)}={\Theta}_{(3)}-i_{\xi}*\mathcal{L}=-d* J_{(2)},
\end{equation}
where $i_{\xi}$ is a contraction of the vector $\xi^{\mu}$ on the first index of $*\mathcal{L}$. The above allows to define a $2-$ form ${Q}_{(2)}=*{J}_{(2)}$ such that $J_{(3)}=d Q_{(2)}$. Finally, taking $\xi^{\mu}$ as a time-like Killing vector that is null on the event horizon, the variation of the Hamiltonian reads
\begin{equation}
\displaystyle{\delta \mathcal{H}=\delta \int_{\mathcal{C}} J_{(3)} -\int_{\mathcal{C}} d \left(i_{\xi} \Theta_{(3)}\right)= \int_{\Sigma^{(2)}}}\left(\delta {Q}_{(2)}-i_{\xi} {\Theta}_{(3)}\right),
\end{equation}
where $\mathcal{C}$ and $\Sigma^{(2)}$ are a Cauchy Surface and its boundary respectively, which has two components, located at infinity ($ \mathcal{H}_{\infty}$) and at the horizon ($\mathcal{H}_{+}$). According to Wald formalism, the first law of black holes thermodynamics
\begin{equation}
d\mathcal{M}= T d\mathcal{S}_{W},\label{eq:first-law}
\end{equation}
is a consequence of $\delta \mathcal{H}_{\infty}=\delta \mathcal{H}_{+}$.
With these ingredients, computing the respective variation of the solution (\ref{eq:mastereq})-(\ref{11}), at the infinity
\begin{equation}
\delta \mathcal{H}_{\infty}=2 \mathcal{Z}_1  \Omega_{2,\epsilon} \delta M \Rightarrow \mathcal{H}_{\infty} = 2 \mathcal{Z}_1  \Omega_{2,\epsilon} M,
\end{equation}
being consistent with the expression for the mass computed previously in (\ref{mass-d4}). While that at the horizon with the Hawking temperature (\ref{1t1})
\begin{equation}
\delta \mathcal{H}_{+}= \left(\epsilon+\frac{3 r_h}{L^2}\right) \mathcal{Z}_1  \Omega_{2,\epsilon} \delta r_h= T \,\delta\left(4  \pi r_h^2\Omega_{2,\epsilon} \mathcal{Z}_1 \right),
\end{equation}
obtaining the entropy of the solution, which reads
\begin{equation}
\mathcal{S}_{W}=4 \pi r_h^2 \, \Omega_{2,\epsilon} \mathcal{Z}_1.\label{eq:sw}    
\end{equation}
It is worth pointing out that for the planar situation ($\epsilon=0$), besides the fulfills of the first law (\ref{eq:first-law}), a four-dimensional Smarr relation \cite{Smarr:1972kt}
\begin{equation}\label{eq:smarr}
\mathcal{M}= \frac{2}{3}\,T \mathcal{S}_{W},   
\end{equation}
 is satisfied. 

With these thermodynamic quantities, in particular the mass $\mathcal{M}$, we are in a position to analyze the NG term, which will be explored in the following section.

\section{Evaluating on the NG action}\label{v3}

In this section, we present the NG action to evaluating to the probe string \cite{Santos:2020xox,Nagasaki:2017kqe,Nagasaki:2018csh}, which is given by
\begin{eqnarray}
S_{NG}=-T_{s}\int{d\sigma\sqrt{-detg_{ind}}},\label{NG1}
\end{eqnarray}
where $T_{s}$ is the fundamental string tension. Adding the Wilson loop implies the insertion of a fundamental string whose world-sheet has a limit in this loop, computing the NG action of this fundamental string. Furthermore, we can calculate the derivative with respect to the time of the NG action,  obtained by integrating the square root of the determinant of the induced metric. In order to study the evolution of the complexity, we draw the Penrose diagram of the causal structure of the hairy black holes given by (\ref{eq:mastereq})-(\ref{11}). For this, to describe the null sheets bounding of the WdW path, it is more convenient to define the tortoise coordinate
\begin{equation}
r^{*}=\int{\frac{dr}{f(r)}},\quad r^{*}_{\infty}=\lim_{r\to\infty}r^{*}(r),\label{NG2}
\end{equation}
where we can define the Eddington-Finkelstein coordinates $u=t+r^{*}$ and $v=t-r^{*}$, which describe outgoing and ingoing null rays respectively. According to \cite{Abad:2017cgl}, the causal structure of the black holes  (\ref{10}) is illustrated by the Penrose diagram in the Figure \ref{pr}, where we are interested in the time dependence of the complexity and therefore in the time dependence of the gravitational action evaluated on this path, as the boundary time increases.
\begin{figure}[!ht]
\begin{center}
\includegraphics[scale=0.7]{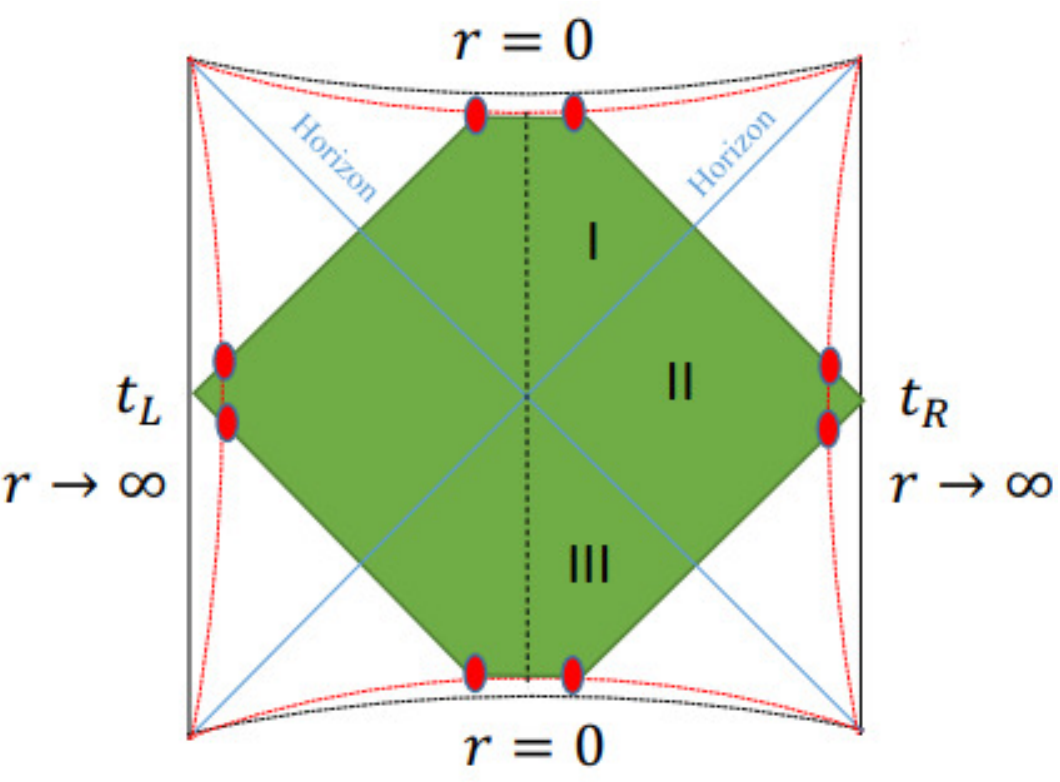}
\includegraphics[scale=0.7]{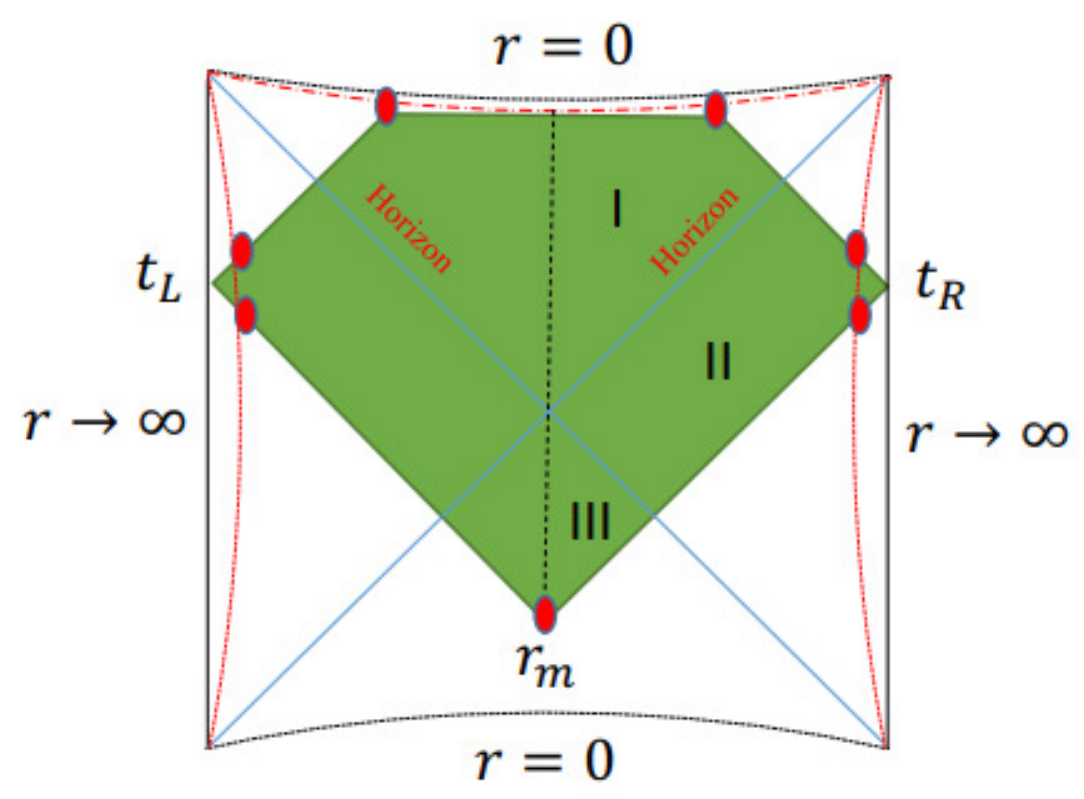}
\caption{Penrose diagram of an eternal black hole. In the WdW patch, this black hole moves forward in time in a symmetric way ($t_{L}=t_{R}$). For both cases, we can see that Region I is the zone behind the future horizon; Region II corresponds to the sector outside both horizons, and Region III is located behind the past horizon.}
\label{pr}
\end{center}
\end{figure}

Now, we will present the effect of the string moving on this spacetime geometry, computing the induced metric using the parameters $\tau$ and $\sigma$ in the world-sheet of the fundamental string. These parameters are given as follows
\begin{eqnarray}
t=\tau,\quad r=\sigma,\quad \varphi=\omega\tau+\xi(\sigma),\label{NG3}
\end{eqnarray} 
where $\omega$ is a constant angular velocity and $\xi(\sigma)$ is a function that determines the string's shape. Considering $d\Omega^{2}_{2,\epsilon}=d \varphi^{2}$, we have that the metric induced in the world-sheet is given by
\begin{eqnarray}
&&ds^{2}_{ind}=H(\sigma)d\tau^{2}+G(\sigma)d\sigma^{2}+2F(\sigma)d\tau d\sigma\label{NG4},\nonumber\\
&&H(\sigma)=-f(\sigma)+\sigma^{2}\omega^{2},\nonumber\\
&&G(\sigma)=\frac{1}{f(\sigma)}+\sigma^{2}\big(\xi' (\sigma)\big)^2,\nonumber\\
&&F(\sigma)=\sigma^{2}\omega\xi^{'}(\sigma),\nonumber\\
&&f(\sigma)=\epsilon+\frac{\sigma^{2}}{L^{2}}-\frac{M}{\sigma},\qquad M=\frac{\mathcal{M}}{ 2 \mathcal{Z}_1 \Omega_{2,\epsilon}},\label{NG5}
\end{eqnarray}
with $L^2$ given previously in (\ref{eq:L2}). \textcolor{black}{Together with the above, we can calculate the time derivative of the NG action (\ref{NG1}), this is obtained through the integration of the square root of the determinant of the induced metric ({\ref{NG4}}), providing by}
\begin{eqnarray}
\frac{dS_{NG}}{dt}=T_{s}\int^{r_{h}}_{0}{d\sigma\sqrt{1-\frac{\sigma^{2}\omega^{2}}{f(\sigma)}+\sigma^{2}(\xi')^{2}f(\sigma)}},\label{NG6}
\end{eqnarray}
where the Lagrangian is given by
\begin{eqnarray}
\mathcal{L}=T_{s}\sqrt{1-\frac{\sigma^{2}\omega^{2}}{f(\sigma)}+\sigma^{2}(\xi')^{2}f(\sigma)},\label{NG7}
\end{eqnarray}
which has the following Euler-Lagrange equation
\begin{eqnarray}
\frac{d}{d\sigma}\left(\frac{\partial\mathcal{L}}{\partial\xi^{'}(\sigma)}\right)-\frac{\partial\mathcal{L}}{\partial\xi(\sigma)}=0.\label{NG8}
\end{eqnarray}
As the Lagrangian is a function of $\mathcal{L}=\mathcal{L}(\sigma,\xi^{'})$, we have that $\partial\mathcal{L}/\partial\xi(\sigma)=0$. Through the equation of motion (\ref{NG8}), we have
\begin{eqnarray}
\displaystyle{\frac{d}{d\sigma}\left(\frac{\partial\mathcal{L}}{\partial\xi^{'}(\sigma)}\right)=\frac{d}{d\sigma}\left(\frac{\sigma^{2}\xi^{'}(\sigma)f(\sigma)}{\sqrt{1-\frac{\sigma^{2}\omega^{2}}{f(\sigma)}+\sigma^{2}(\xi')^{2}f(\sigma)}}\right)=0.}\label{NG10}
\end{eqnarray}
Then, the equation (\ref{NG10}) can be written as
\begin{eqnarray}
c_{\xi}:=\frac{\sigma^{2}\xi^{'}(\sigma)f(\sigma)}{\sqrt{1-\frac{\sigma^{2}\omega^{2}}{f(\sigma)}+\sigma^{2}(\xi')^{2}f(\sigma)}}=\frac{\sigma^{2}\xi^{'}(\sigma)f(\sigma)}{(\mathcal{L}/T_{s})},\label{NG11}
\end{eqnarray}
where the explicit expression for $\xi^{'}(\sigma)$, after some algebraic manipulations, reads
\begin{eqnarray}
\xi^{'}(\sigma)=\frac{c_{\xi}}{\sigma f(\sigma)}\sqrt{\frac{f(\sigma)-\sigma^{2}\omega^{2}}{\sigma^{2}f(\sigma)-c^{2}_{\xi}}}.\label{NG12}
\end{eqnarray}
For the rational function inside the square root from (\ref{NG12}), the denominator must be negative when the numerator $f(\sigma)-\sigma^{2}\omega^{2}$ becomes negative, and they become zero coincidentally.  Using this condition, we can find the $c_{\xi}$ constant which reads
\begin{eqnarray}
&&c_{\xi}=\sigma^{2}_{H}\omega\label{NG13},\\
&&\sigma_{H}=-\frac{(2/3)^{1/3}\epsilon L}{\left[9(1-v^{2})(M/L)+\sqrt{3}\sqrt{27(1-v^{2})^{4}(M/L)^{2}-4\epsilon^{3}(1-v^{2})^{3}}\right]^{1/3}}+\nonumber\\&&+\frac{L\left[9(1-v^{2})(M/L)+\sqrt{3}\sqrt{27(1-v^{2})^{4}(M/L)^{2}-4\epsilon^{3}(1-v^{2})^{3}}\right]^{1/3}}{2^{1/3}(3)^{2/3}},
\end{eqnarray}
where $\sigma=\sigma_{H}$ is the solution for numerator of the square root in equation (\ref{NG12}), and $v=\omega L$ is the angular velocity. We can observe that $\mathcal{L}=T_{s}\sigma\xi^{'}f(\sigma)/c_{\xi}$, and we can write
\begin{eqnarray}
\frac{dS_{NG}}{T_{s}dt}&=&\displaystyle{\int^{r_{h}}_{0} d\sigma \sqrt{ {\frac {\sigma\, \left[  \left( {\sigma}^{2}+\sigma \sigma_{H}+
\sigma_{H}^{2} \right)  \left( 1-{v}^{2} \right) +\epsilon\,{L}^{2}
 \right] }{{\sigma}^{3}+{\sigma}^{2}\sigma_{H}+ \left(\sigma_{H}^{2}+\epsilon\,{L}^{2} \right) \sigma+\sigma_{H}^{3}{v}^{2}}}
}},\label{NG14}
\end{eqnarray}
\begin{eqnarray}
r_{h}&=&\displaystyle{-\frac{(2/3)^{1/3}\epsilon L}{\left[9(M/L)+\sqrt{3}\sqrt{27(M/L)^{2}+4\epsilon^{3}}\right]^{1/3}}}\nonumber\\
&+&\displaystyle{\frac{L\left[9(M/L)+\sqrt{3}\sqrt{27(M/L)^{2}+4\epsilon^{3}}\right]^{1/3}}{2^{1/3}(3)^{2/3}}}.\label{NG15}
\end{eqnarray}

{As we can see, this integral is similar to the solutions explored in \cite{Nagasaki:2018csh}. Nevertheless, is interesting to note that in (\ref{NG14}) the action growth depends on the mass of the black hole (\ref{mass-d4}) as well as the angular velocity (which depends of $L$ from (\ref{eq:L2})), where the functions $\mathcal{Z}_{1}(X)$ and $Z(X)$ are \textcolor{black}{present}. This shows the importance  of the coupling functions given by the DHOST theory (\ref{action})-(\ref{lagrangian}). Together with the above, as in \cite{Nagasaki:2018csh}, we perform numerical procedures to solve the integral (\ref{NG14})-(\ref{NG15}), being reflected in the Figures \ref{p}-\ref{p2} for the three topologies, with $\epsilon=1$, $\epsilon=0$, and $\epsilon=-1$ respectively. }

\begin{figure}[!ht]
\begin{center}
\includegraphics[scale=0.163]{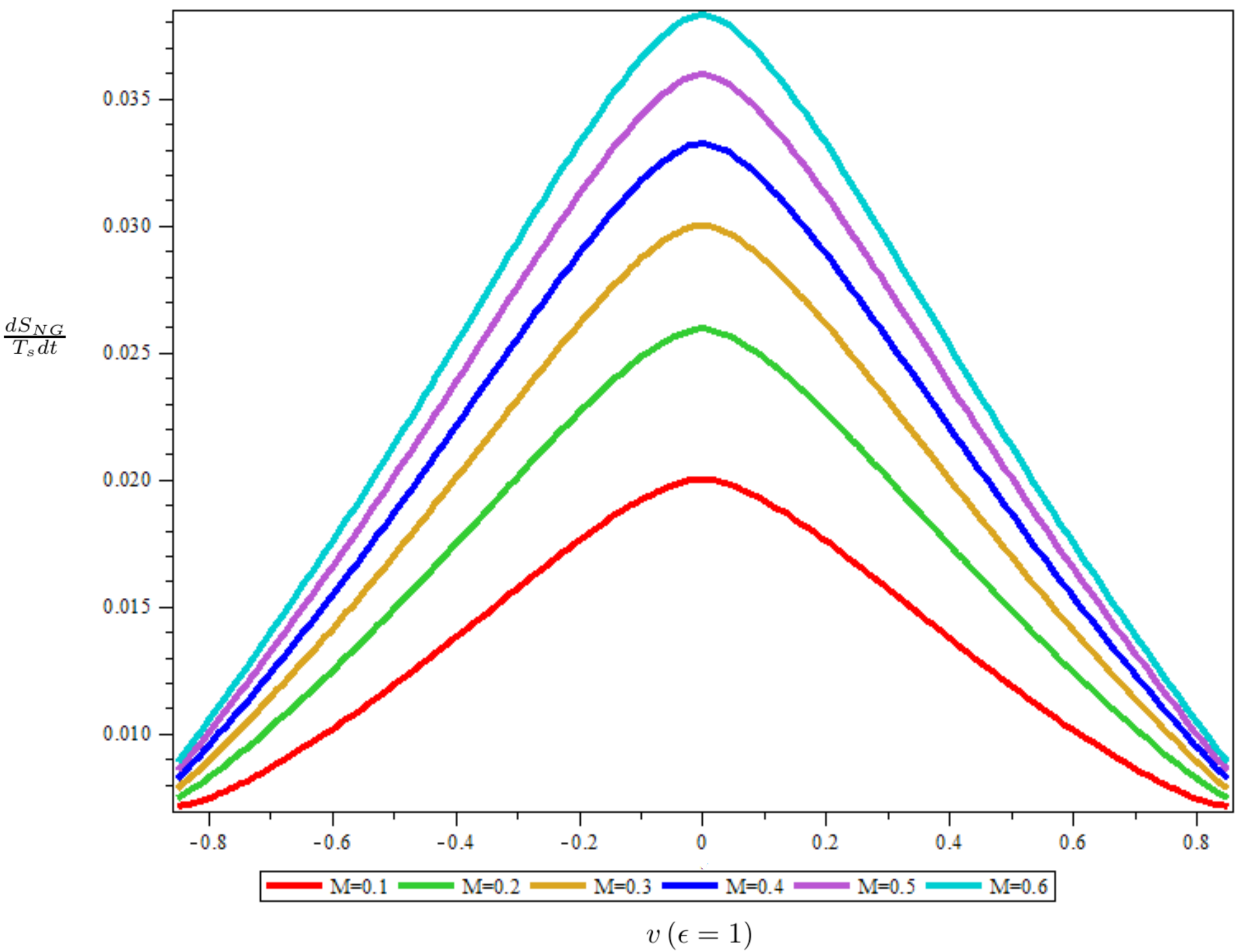} 
\includegraphics[scale=0.163]{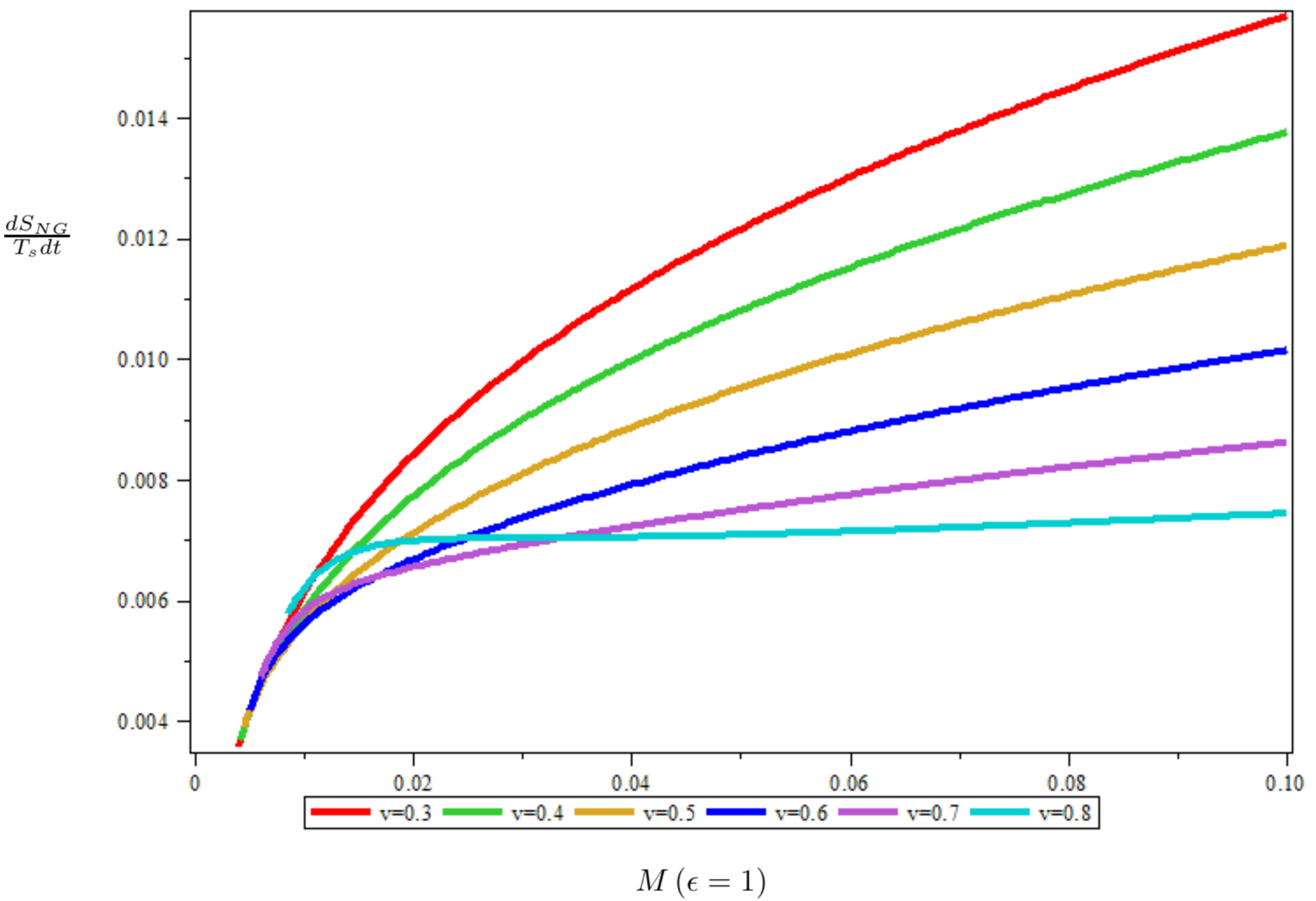} 
\caption{Graphics of the action growth versus angular velocity $v$ and the mass dependence $M={\mathcal{M}}/( 2 \mathcal{Z}_1 \Omega_{2,1})$ with $\epsilon=1$. \textcolor{black}{Top panel}: Action growth for the hairy Black hole (\ref{10})-(\ref{11})  and angular velocity dependence $v$ for the values: $M=0.1$, $M=0.2$, $M=0.3$,  $M=0.4$, $M=0.5$ $M=0.6$. \textcolor{black}{Bottom panel:} Action growth for the hairy Black hole (\ref{10})-(\ref{11}) with  for the values: $v=0.3$, $v=0.4$, $v=0.5$, $v=0.6$, $v=0.7$,$v=0.8$. For both cases, we impose $L=0.01$ and $\Omega_{2,1}=1$.}
\label{p}
\end{center}
\end{figure}

\begin{figure}[!ht]
\begin{center}
\includegraphics[scale=0.163]{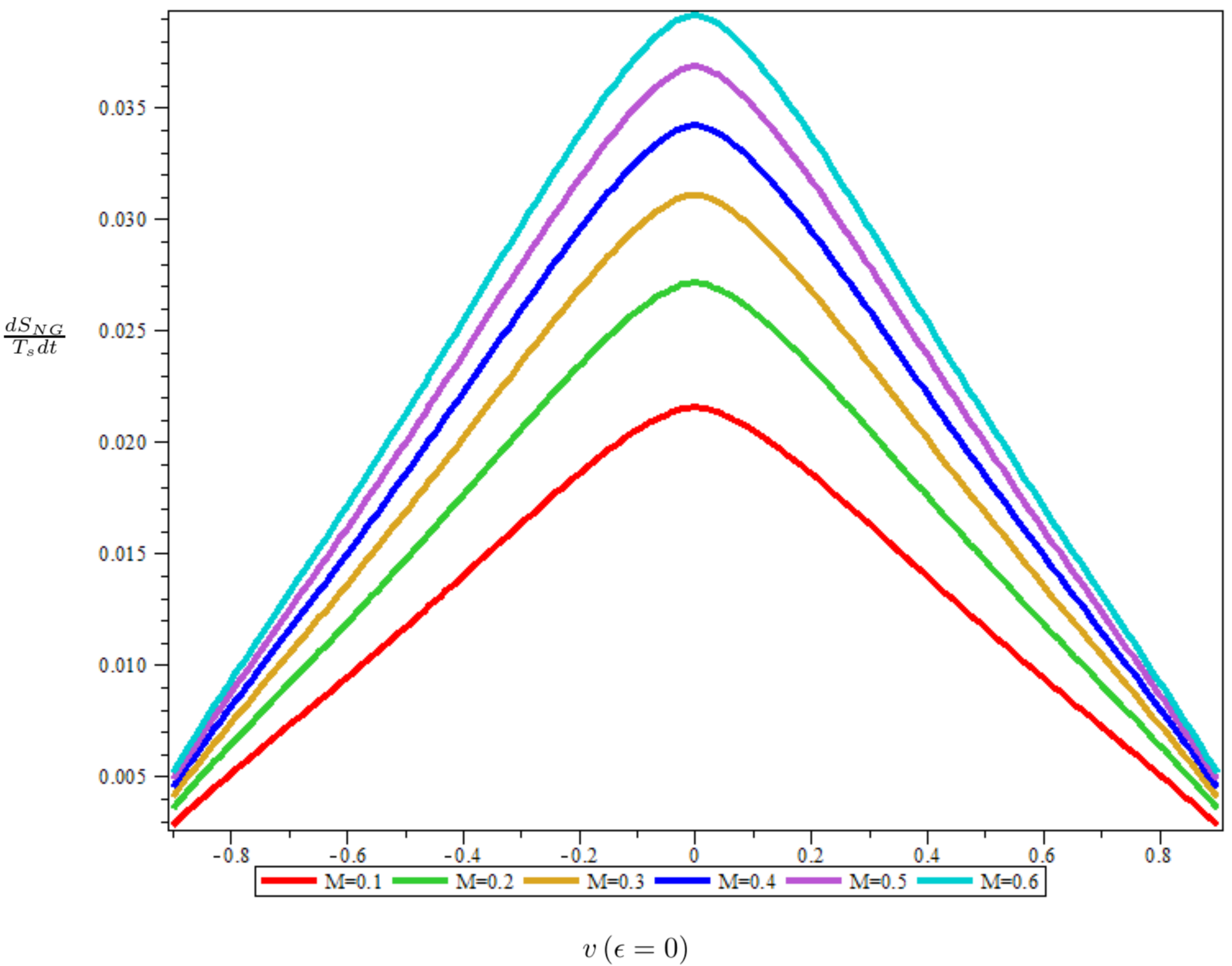}
\includegraphics[scale=0.163]{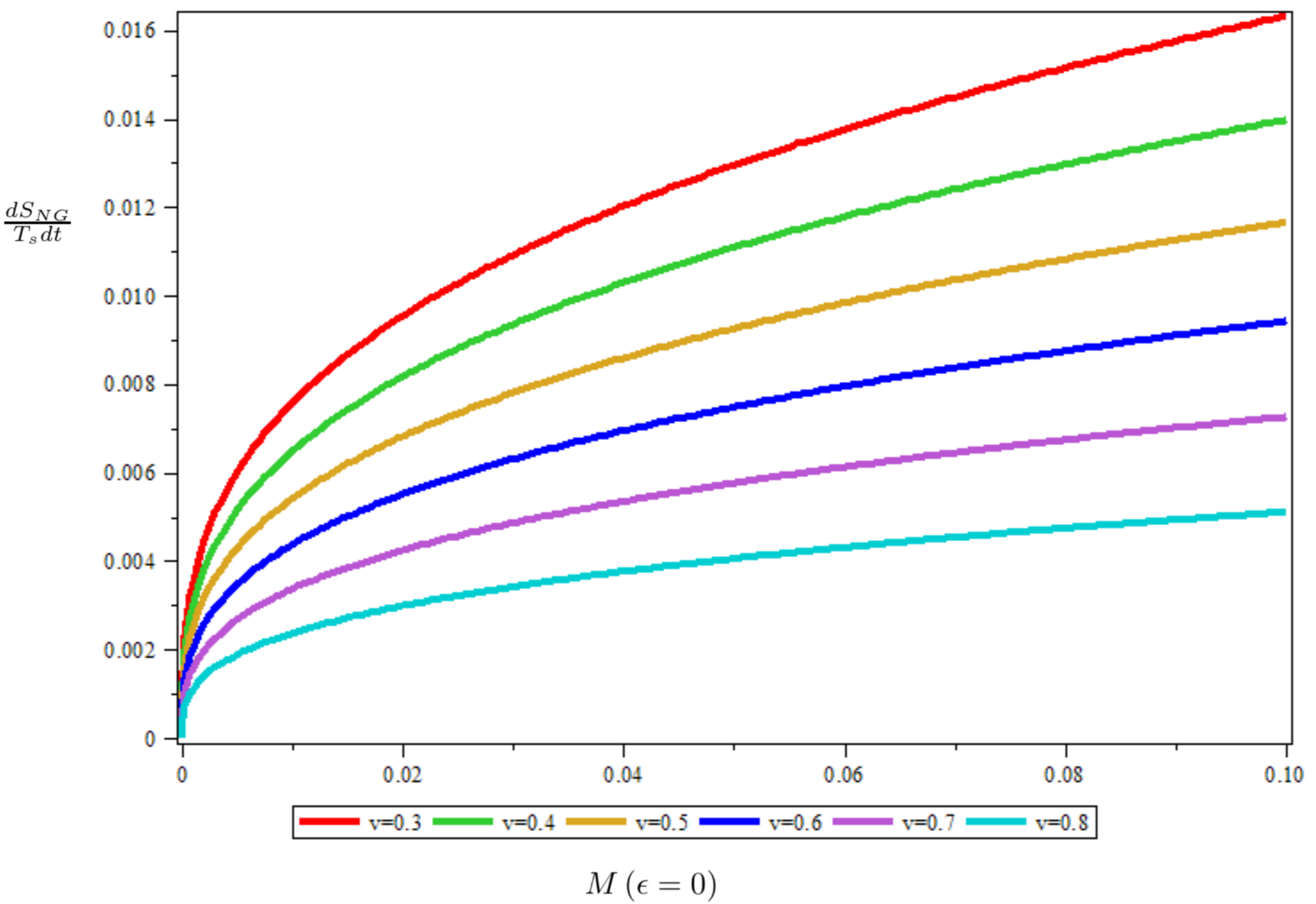}
\caption{Graphics of the action growth versus angular velocity $v$ and the mass dependence $M={\mathcal{M}}/( 2 \mathcal{Z}_1 \Omega_{2,0})$ with $\epsilon=0$. \textcolor{black}{Top panel}: Action growth for the hairy Black hole (\ref{10})-(\ref{11})  and angular velocity dependence $v$ for the values: $M=0.1$, $M=0.2$, $M=0.3$,  $M=0.4$, $M=0.5$ $M=0.6$. \textcolor{black}{Bottom panel:} Action growth for the hairy Black hole (\ref{10})-(\ref{11}) with  for the values: $v=0.3$, $v=0.4$, $v=0.5$, $v=0.6$, $v=0.7$,$v=0.8$. For both cases, we impose $L=0.01$ and $\Omega_{2,0}=1$.}
\label{p1}
\end{center}
\end{figure}

\begin{figure}[!ht]
\begin{center}
\includegraphics[scale=0.16]{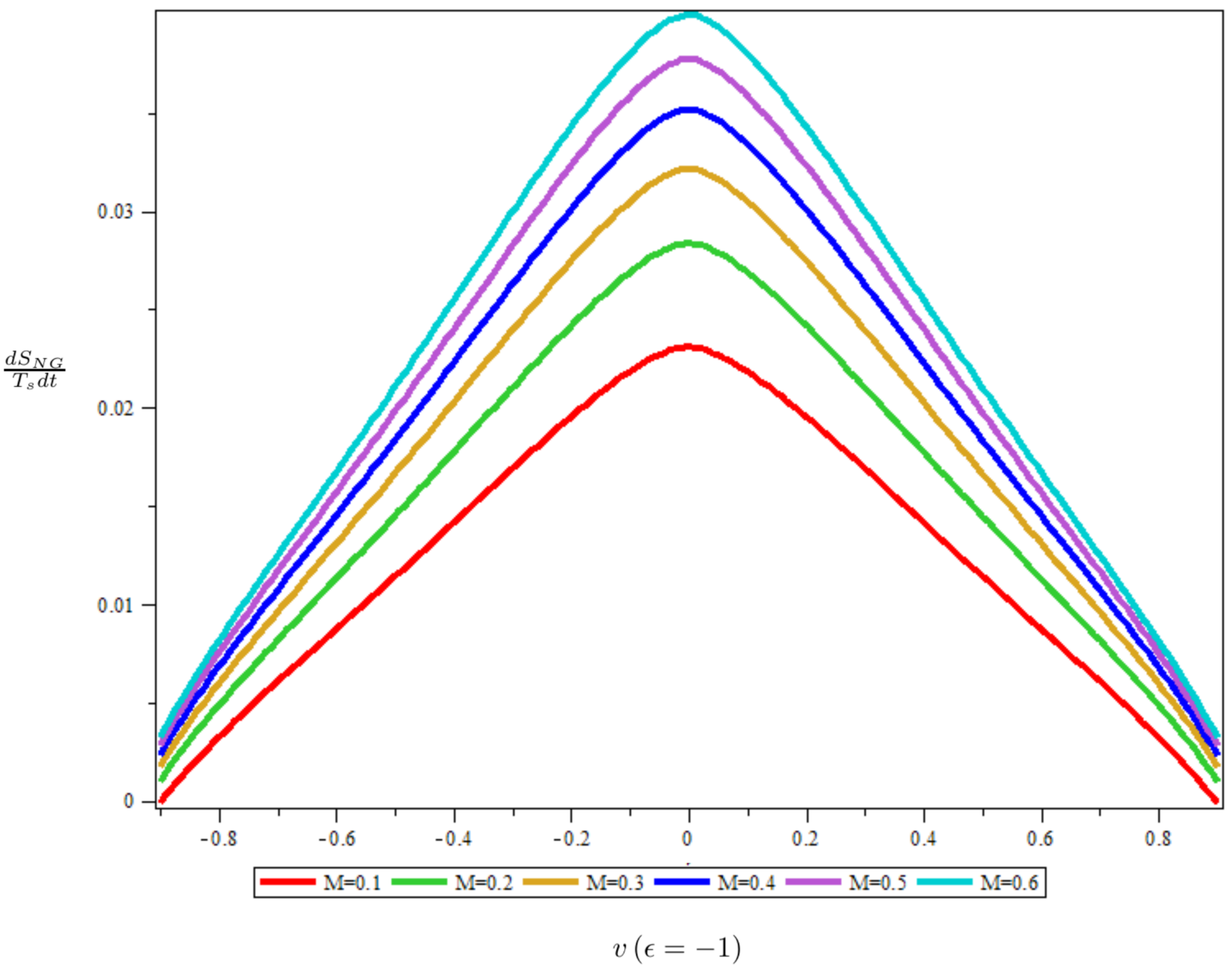}
\includegraphics[scale=0.16]{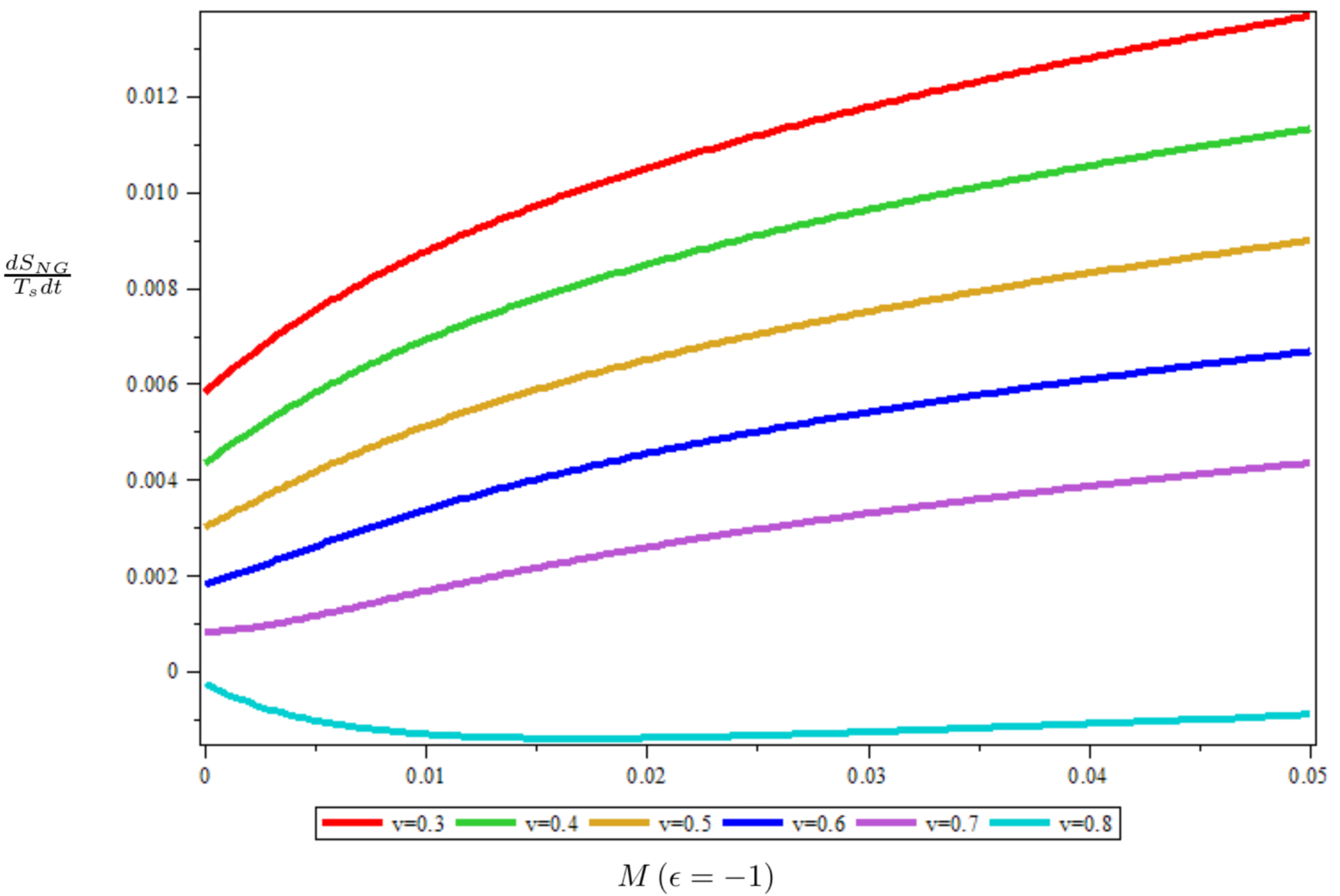}
\caption{Graphics of the action growth versus angular velocity $v$ and the mass dependence $M={\mathcal{M}}/( 2 \mathcal{Z}_1 \Omega_{2,-1})$ with $\epsilon=-1$. \textcolor{black}{Top panel}: Action growth for the hairy Black hole (\ref{10})-(\ref{11})  with angular velocity dependence $v$ for the values: $M=0.1$, $M=0.2$, $M=0.3$,  $M=0.4$, $M=0.5$ $M=0.6$. \textcolor{black}{Bottom panel:} Action growth for the hairy Black hole (\ref{10})-(\ref{11}) with  for the values: $v=0.3$, $v=0.4$, $v=0.5$, $v=0.6$, $v=0.7$,$v=0.8$. For both cases, we impose $L=0.01$ and $\Omega_{2,-1}=1$.}
\label{p2}
\end{center}
\end{figure}

According to Figure \ref{p}, from the \textcolor{black}{top panel}, we can see that the growth of the action is larger as the mass increases and the string moves slower, while that when the string moves faster, the effect becomes less, showing us the effect of the drag force. In addition, from the \textcolor{black}{bottom panel} we have that when the angular velocity is small, the growth action is an increasing function of the mass. Nevertheless, for some special value of $v$ near to one, becomes decreasing. 

For the planar case, as shown in Figure \ref{p1}, we have that the action growth versus angular velocity (\textcolor{black}{top panel})  resemble the spherical case.  In addition, the action growth versus  $M$ (\textcolor{black}{bottom panel}) provides us that always is an increasing function.

Figure \ref{p2} shows that for the hyperbolic case we have that the action growth for regions of great mass corresponds to the region where the string moves more slowly (\textcolor{black}{top panel}). Nevertheless,  from the \textcolor{black}{bottom panel}, we note that exists some values of the angular velocity, near to one, where the action growth is simultaneously a decreasing and then an increasing function of $M$, unlike the spherical or planar situation. We can also note that for this case, the causal structure of the black hole has two horizons, one external and one internal. Nevertheless, as was explained in Section \ref{v2}, we are focused on the case with $r_{h}>L$, where the mass $\mathcal{M}$ is always positive.

{\color{black}In resume, and according to the CA conjecture where the action growth is equal to the complexity growth, such growth of the complexity for the three cases are controlled by coupling functions present in the DHOST theory (\ref{action})-(\ref{Li}), where  we can see that this growth is
larger as the mass increases and the string moves slower (top panel of Figures \ref{p}-\ref{p2}) . This result is consistent concerning the analysis performed in discussions of \cite{Nagasaki:2018csh}, where now for the planar or spherical case ($\epsilon\in \{0,1\}$), we have that the growth of the complexity is an increasing function of the mass when the angular velocity is small, unlike the hyperbolical situation ($\epsilon=-1$), where we can find some values where the action growth is simultaneously a decreasing and then an increasing function of $M$, as shown in the bottom panel of  Figures \ref{p}-\ref{p2}. Additionally, in general,  we have that the velocity dependence of the stationary string gives the maximal complexity growth, where the growth of the complexity decreases as the probe string moves faster. }  

\textcolor{black}{On the other hand, althought }the above analysis is similar to the cases studied in  \cite{Nagasaki:2018csh,Nagasaki:2017kqe}, in Figures \ref{p}-\ref{p2} we impose that the radius $L$ takes a particular value, but this parameter as well as the mass depend on the functions $Z$ and $\mathcal{Z}_1$. In fact, if we suppose that for the spherical case the mass  $M={\mathcal{M}}/( 2 \mathcal{Z}_1 \Omega_{2,1})$ is fixed, yet we can change the values of $L$ from (\ref{eq:L2}) with a suitable election of the coupling functions satisfy (\ref{eq:mastereq}), given in the Figure \ref{pc}. The above shows us the importance of the coupling functions from the DHOST theory (\ref{action})-(\ref{Li}).

\begin{figure}[!ht]
\begin{center}
\includegraphics[scale=0.165]{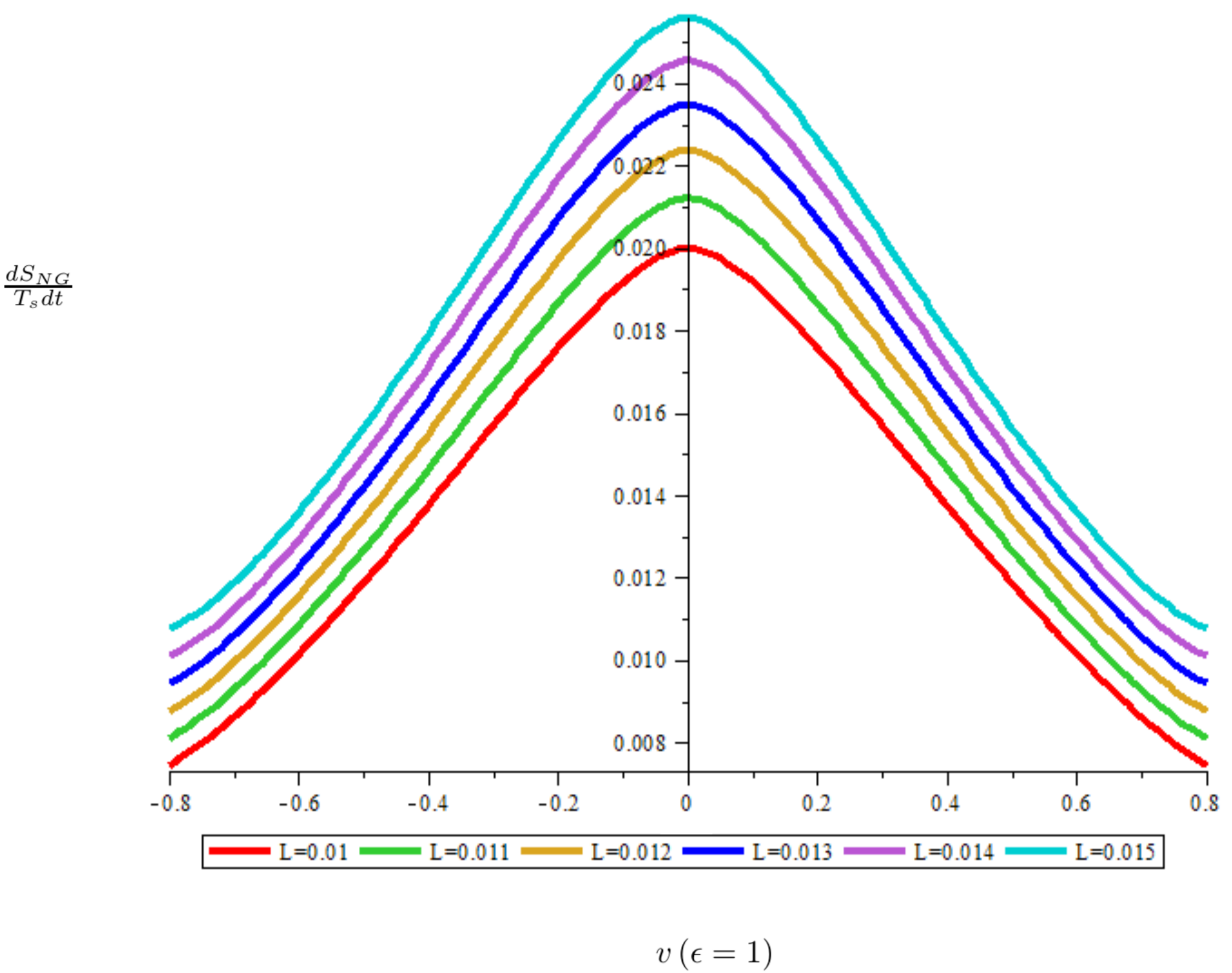}
\caption{Graphics of the action growth versus angular velocity $v$ and a fixed mass with $\epsilon=1$, where $M=0.1$ and for values of the radius $L$:  $L=0.01$, $L=0.012$, $L=0.013$,  $L=0.014$, $L=0.015$.}
\label{pc}
\end{center}
\end{figure}

In the following section, we will explore the shear viscosity for the planar hairy configuration  (this is, the solution (\ref{eq:mastereq})-(\ref{11}) with $\epsilon=0$), where the thermodynamical quantity $\mathcal{S}_{W}$ computed previously in (\ref{eq:sw}) has a providential role.

\section{The viscosity/entropy density ratio}
\label{viscosity}

As was shown in the introduction, AdS-planar black hole configurations allows us to study the viscosity/\textcolor{black}{entropy density}$(\eta/s)$ ratio. Just to clarify, in the following computations we impose $\epsilon=0$ from the thermodynamical quantities and the solution found in (\ref{eq:mastereq})-(\ref{11}). As a first step, we perform a transverse and traceless perturbation on the metric (\ref{7}) with $h=f$, which reads 
\begin{equation}
ds^2=-f(r) dt^2+\frac{dr^2}{f(r)}+2 r^2 \Psi(t,r) dx_1 dx_2+ r^2 \left(d x_{1}^2+d x_{2}^2 \right),
\end{equation}
 where with the Ansatz 
\begin{equation}
\Psi(t,r) =\gamma t + h_{x_1 x_2}(r),
\end{equation}
with $\gamma$ a constant identified as the gradient of the fluid velocity along the $x_1$ direction \cite{Fan:2018qnt}, yields the following linearized equations for $h_{x_1 x_2}$:
\begin{eqnarray}\label{linear}
\left[\mathcal{Z}_{1} r^{2} f (h_{x_1 x_2})'\right]'=0.
\end{eqnarray}
According to the Wald formalism \cite{Iyer:1994ys,Wald:1993nt} together with the method \cite{Fan:2018qnt}, the shear stress is associated to the current
\begin{equation}
{\mathcal{J}}^{x_2}= \sqrt{-g} {Q}^{r x_2}= \mathcal{Z}_{1}\,r^{2} f (h_{x_1 x_2})', 
\end{equation}\\
which is conserved due to $({\mathcal{J}^{x_2}})'$ is the linear equation (\ref{linear}), constructed with the Noether potential $ {\mathcal{Q}}^{\mu \nu}$ given previously in (\ref{eq:noether}) and a space-like Killing vector $\partial_{x_1}=\xi^{\mu}\partial_{\mu}$. Imposing the ingoing horizon boundary condition
\begin{equation}
h_{x_1 x_2} = \gamma \sqrt{\frac{G}{\mathcal{Z}_1}}\,\frac{\log (r-r_h)}{4 \pi T}+\cdots,
\end{equation}
as well as a Taylor expansion in the near horizon region $r_h$
\begin{equation}
h=f=4\pi T (r-r_h)+\cdots,
\end{equation}
where $T$ is the Hawking temperature (\ref{1t1}), we have:
\begin{equation}
\eta = \gamma  \mathcal{Z}_1 \, \sqrt{\frac{G}{\mathcal{Z}_1}}\, r_h^{2}=   \frac{1}{4 \pi} \sqrt{\frac{G}{\mathcal{Z}_1}} \gamma s, 
\end{equation}
with $s$ corresponding to the \textcolor{black}{entropy density} given by
\begin{equation}
s=\frac{\mathcal{S}_{W}}{\Omega_{2,0}}=4 \pi r_h^{2} \mathcal{Z}_1.
\end{equation}
Finally, the viscosity/\textcolor{black}{entropy density} ratio takes the form
\begin{equation}\label{vis/den}
\textcolor{black}{\frac{\eta}{s}=\frac{1}{4 \pi}  \sqrt{\frac{G}{\mathcal{Z}_1}}=\frac{1}{4 \pi}  \sqrt{\frac{G}{G+X A_2}.}}
\end{equation}
Note that, as the examples computed in \cite{Buchel:2003tz,Buchel:2004qq,Benincasa:2006fu,Landsteiner:2007bd}, there is not a presence of the location of the event horizon $r_h$ on the $\eta/s$ ratio  (\ref{vis/den}), where for a suitable election of the kinetic term, the \textcolor{black}{ratio} goes back to the universal value $1/(4 \pi)$. For example, in addition to the condition (\ref{eq:mastereq}), it is possible to find a $X$ such that $G=\mathcal{Z}_{1}$ (see for example the case $G=\kappa$, $Z=-2\Lambda -\gamma X$ and $A_{2}=0$), where the KKS bound is satisfied. In fact, the Einstein-Hilbert action together with a cosmological constant is recovered from this case.   Nevertheless, in (\ref{vis/den})  \textcolor{black}{appears a contribution depending on the functions $G$ and ${\mathcal{Z}_{1}}$, where the four dimensional viscosity/entropy density ratio for special cases studied in \cite{Feng:2015oea} and \cite{Bravo-Gaete:2021hlc} are recovered, violating the KSS bound.
Together with the above,  the $\eta/s$ ratio (\ref{vis/den})  allows us to construct examples where the kinetic term $X$ satisfies (\ref{eq:mastereq})-(\ref{eq:mastereq2}) and}
\begin{equation}
\textcolor{black}{0<\frac{G}{{\mathcal{Z}_{1}}}=\frac{G}{G+X A_2}<1,}
\end{equation}
with a suitable choice of coupling functions. For example, for
\begin{equation}
G(X)=\kappa + k X^{j}, \qquad A_2 =G_{X}= k j X^{j-1},
\end{equation}
where $\kappa, X, k $ and $j$ are positive constants, while that $Z$ and $\mathcal{Z}_{2}$ satisfy (\ref{eq:mastereq}), we have
\begin{equation}
0<\frac{G}{{\mathcal{Z}_{1}}}=\frac{\kappa + k X^{j}}{\kappa + (1+j)k X^{j} }<1,
\end{equation}
implying that
\begin{equation}
0<\frac{\eta}{s}<\frac{1}{4\pi}.
\end{equation}
Just for completeness, the viscosity/ \textcolor{black}{entropy density} ratio (\ref{vis/den}) also can be found following the steps described in \cite{Son:2002sd,Brigante:2007nu}.

\section{Conclusions and discussions}\label{v4}

The aims of the present work were, first, to show that generalized scalar-tensor theories present in (\ref{action})-(\ref{lagrangian}) admit hairy black holes. Indeed, for this model, the solution resembles  the four-dimensional metric with the presence of an effective cosmological constant and a non-trivial expression for the scalar field, with three different topologies, where we are supposing that the kinetic term is a constant. With this result, we exhibit a variety of thermodynamical behaviors, via the quasilocal method and Wald formalism, showing that the first law of black holes thermodynamics holds (\ref{eq:first-law}), together with a Smarr relation for the planar situation (\ref{eq:smarr}).

With the above, our second motivation was to compute the Nambu-Goto action, obtaining the relation between the action growth and the angular velocity as well as the mass of these solutions, showing different behaviors depending on the topology of the event horizon, presented in Figures \ref{p}-\ref{p2},  where according to CA conjecture, these situations represent the growth of the black-hole complexity. Additionally,  we note the importance of these analyses of the coupling functions $\mathcal{Z}_1(X)$ and $Z(X)$, present in this generalized scalar-tensor model. 

As a final aim, for the planar case, we explored the viscosity/\textcolor{black}{entropy density} ratio, constructing a conserved charge via the Wald formalism, and a suitable election of the Killing vector, obtaining that this ratio depends on the coupling functions from the theory ($G$ and $\mathcal{Z}_1$). Here, some particular cases where the KSS bound is not fulfilled were studied. To our knowledge, together with \cite{Brito:2019ose},\cite{Feng:2015oea}, our work provides a new example of the violation of the  $\eta/s$ ratio whose Lagrangian is at most linear in curvature tensor.

Further works are, naturally, to analyze hairy black holes where the kinetic term $X$ is not a constant, higher dimensional configurations and even to study the CA conjecture by using the procedure elaborated in \cite{Lehner:2016vdi}, giving a deep discussion of the boundary terms required in the action functional, having an attention to the case of null boundary segments, needing the introduction to associated joints contributions.

\acknowledgments

FS would like to thank CNPq and CAPES for partial financial support. FS also would like to thank Komeil Babaei Velni, Yue-Zhou Li, Dmitry Ageev, and Kumar Shwetketu Virbhadra for valuable comments and discussions. The authors thank the Referee for the commentaries and suggestions to improve the paper.

\section{Appendix}
\label{sec:app}
Here we report the expressions for ${\cal{G}}^{Z}_{\mu\nu}
,{\cal{G}}^{G}_{\mu\nu}$, the ${\cal{G}}^{(i)}_{\mu\nu}$'s and $\mathcal{J}^{\mu}$ present in the equations of motions (\ref{eq:emotion1})-(\ref{eq:emotion2})
\begin{align*}
{\cal{G}}^{Z}_{\mu\nu}&=-\frac{1}{2} Z(X) g_{\mu\nu}+Z_{X} \phi_{\mu}
\phi_{\nu}, \\ \\
{\cal{G}}^{G}_{\mu\nu}&= G  G_{\mu\nu}+G_{X} R \phi_{\mu} \phi_{\nu} -\nabla_{\nu} \nabla_{\mu} G  \\
&+g_{\mu\nu} \nabla_{\lambda}\nabla^{\lambda} G,
\\ \\
{\cal{G}}^{(2)}_{\mu\nu}&=-\phi_{\mu}\,(A_{2X} \nabla_{\nu}X)\,\Box \phi-(A_{2X} \nabla_{\mu} X ) \phi_{\nu} \,\Box \phi\\
&-A_2  \phi_{\nu \mu} \Box \phi -\phi_{\nu \mu} \phi_{\lambda} ( A_{2X} \nabla^{\lambda}
X)
\\
&+\phi_{\nu} \phi_{\lambda \mu} (A_{2X} \nabla^{\lambda}X)+\phi_{\mu}
\phi_{\lambda\nu} (A_{2X} \nabla^{\lambda} X)\\
&+A_2  R_{\nu\lambda} \phi_{\mu} \phi^{\lambda}+A_2  R_{\mu\lambda} \phi_{\nu}
\phi^{\lambda}\\
&-A_2  \phi_{\lambda\nu\mu} \phi^{\lambda}+\frac{1}{2}A_2  g_{\mu\nu} (\Box \phi)^2\\
&+g_{\mu\nu} \phi_{\lambda}  (A_{2X} \nabla^{\lambda} X) \Box \phi +A_2  g_{\mu\nu}
\phi^{\lambda}  \phi^{\,\,\rho \,}_{\rho \,\,\lambda}
\\
&-A_2  g_{\mu\nu} R_{\lambda\rho} \phi^{\lambda} \phi^{\rho}
+\frac{1}{2} A_2  g_{\mu\nu} \phi_{\rho\lambda} \phi^{\rho\lambda}\\
&+A_{2X} \phi_{\mu} \phi_{\nu}  \big((\Box \phi)^2-\phi_{\lambda\rho}
\phi^{\lambda\rho}\big),
\end{align*}
\begin{align*}
{\cal{G}}^{(3)}_{\mu\nu}&=-\frac{1}{2}A_3  \phi_{\mu} \phi_{\nu} (\Box
\phi)^2-
\frac{1}{2} \phi_{\mu} \phi_{\nu} \phi_{\lambda}  (A_{3X} \nabla^{\lambda}X) \Box \phi\\
&+\frac{1}{2} A_{3}  \phi_{\mu}  \phi_{\lambda\nu} \phi^{\lambda} \Box \phi+
\frac{1}{2} A_{3}  \phi_{\nu} \phi_{\lambda\mu}\phi^{\lambda}
\Box \phi\\
&-\frac{1}{2} A_{3}  \phi_{\mu} \phi_{\nu} \phi^{\lambda} \phi_{\rho\,\, \lambda}^{\,\,
\rho \,} +\frac{1}{2} A_3  R_{\lambda\rho} \phi_{\mu} \phi_{\nu} \phi^{\lambda}
\phi^{\rho}\\
&-\frac{1}{2} \phi_{\mu}  (A_{3X} \nabla_{\nu} X) \phi^{\lambda}  \phi_{\rho\lambda}
\phi^{\rho} -\frac{1}{2} (A_{3X} \nabla_{\mu} X )\phi_{\nu}
\phi^{\lambda} \phi_{\rho\lambda} \phi^{\rho} \\
&-\frac{1}{2}A_3  \phi_{\nu} \phi^{\lambda} \phi_{\rho\lambda\mu}
\phi^{\rho}-\frac{1}{2}A_3
\phi_{\mu} \phi^{\lambda}  \phi_{\rho\lambda\nu} \phi^{\rho}\\
&-A_3  \phi_{\nu} \phi^{\lambda} \phi_{\rho\lambda} \phi^{\rho}_{\,\,\mu}-A_3  \phi_{\mu}
\phi^{\lambda}
\phi_{\rho\lambda} \phi^{\rho}_{\,\ \nu} \\
&+\frac{1}{2} g_{\mu\nu} \phi_{\lambda} (A_{3X} \nabla^{\lambda} X) \phi^{\rho}
\phi_{\sigma\rho}  \phi^{\sigma} +\frac{1}{2} g_{\mu\nu} A_{3}  \phi^{\lambda}  \phi^{\rho}
\phi_{\sigma\rho\lambda} \phi^{\sigma}\\
&+g_{\mu\nu} A_{3}  \phi^{\lambda}  \phi^{\rho} \phi_{\sigma\rho}
\phi^{\sigma}_{\,\,\lambda}+A_{3X} \phi_{\mu}
\phi_{\nu} (\Box \phi) \phi^{\rho} \phi_{\sigma\rho} \phi^{\sigma}, \\ \\
{\cal{G}}^{(4)}_{\mu\nu}&=-A_4  \phi_{\mu} \phi_{\nu} \phi^{\lambda}
\phi_{\rho\,\,\lambda}^{\,\,\rho}+ A_4  \phi_{\lambda\mu} \phi^{\lambda} \phi_{\rho\nu} \phi^{\rho}
\\
&-\phi_{\mu}  \phi_{\nu}  (A_{4X} \nabla^{\lambda} X) \phi_{\rho\lambda} \phi^{\rho} -A_4
\phi_{\mu} \phi_{\nu}
\phi_{\rho\lambda} \phi^{\rho\lambda}\\
&-\frac{1}{2} A_4  g_{\mu\nu} \phi^{\lambda} \phi^{\rho} \phi_{\sigma\rho}
\phi^{\sigma}_{\,\,\lambda} +A_{4X} \phi_{\mu} \phi_{\nu} \phi_{\lambda\rho} \phi^{\lambda}
\phi^{\rho\sigma} \phi_{\sigma},\\ \\
{\cal{G}}^{(5)}_{\mu\nu}&=-A_5  \phi_{\mu} \phi_{\nu} \phi^{\lambda} \phi_{\rho\lambda}
\phi^{\rho} (\Box \phi)
-\phi_{\mu} \phi_{\nu}  \phi_{\lambda} (A_{5X} \nabla^{\lambda} X) \phi^{\rho} \phi_{\sigma\rho} \phi^{\sigma}\\
&+A_5  \phi_{\nu}  \phi_{\lambda\mu} \phi^{\lambda} \phi^{\rho} \phi_{\sigma\rho} \phi^{\sigma}+
A_5  \phi_{\mu} \phi_{\lambda\nu} \phi^{\lambda} \phi^{\rho}
\phi_{\sigma\rho} \phi^{\sigma}\\
&-A_5  \phi_{\mu} \phi_{\nu}  \phi^{\lambda} \phi^{\rho}
\phi_{\sigma\rho\lambda} \phi^{\sigma}-2A_5  \phi_{\mu} \phi_{\nu}  \phi^{\lambda} \phi^{\rho} \phi_{\sigma\rho} \phi^{\sigma}_{\,\,\lambda}\\
&-\frac{1}{2} A_5  g_{\mu\nu} \phi^{\lambda} \phi_{\rho\lambda} \phi^{\rho} \phi^{\sigma}
\phi_{\tau\sigma} \phi^{\tau}+A_{5X} \phi_{\mu}  \phi_{\nu}
 \phi^{\lambda} \phi^{\rho} \phi_{\rho\lambda}
\phi^{\sigma}\phi^{\tau} \phi_{\tau\sigma},
\end{align*}
while that
\begin{eqnarray*}
\mathcal{J}^{\mu}=\mathcal{J}^{\mu}_{Z}+\mathcal{J}^{\mu}_{G}+\sum_{i=2}^{5} \mathcal{J}^{\mu}_{(i)},
\end{eqnarray*}
with
\begin{eqnarray*}
\mathcal{J}^{\mu}_{Z}&=&2 Z_{X} \phi^{\mu}, \\ \\
\mathcal{J}^{\mu}_{G}&=&2 G_{X} R \phi^{\mu}, 
\end{eqnarray*}
\begin{eqnarray*}
\mathcal{J}^{\mu}_{(2)}&=& 2 A_{2X} \phi^{\mu}\left[ (\Box \phi)^{2}-
\phi_{\lambda \rho}\phi^{\lambda \rho}\right]-2\nabla_{\nu} \left[ A_2  \left( g^{\mu \nu}-\phi^{\mu \nu}\right)\right],
\\ \\
\mathcal{J}^{\mu}_{(3)}&=&2 A_{3X} \phi^{\mu}  \, \Box \phi\,  \phi^{\lambda}
\phi_{\lambda \rho} \phi^{\rho}+2 A_3  \,\Box \phi \,
\phi^{\mu}_{\,\,\,\lambda}\phi^{\lambda}\\
&-&\nabla_{\nu} \left[A_3  \big(g^{\mu \nu} \phi^{\lambda} \phi_{\lambda \rho} \phi^{\rho}+\Box \phi \,\phi^{\mu} \phi^{\nu}\big)\right],\\ \\
\mathcal{J}^{\mu}_{(4)}&=&2 A_{4X} \phi^{\mu} \phi^{\sigma} \phi_{\sigma \rho}
\phi^{\rho \lambda}
\phi_{\lambda}+A_4(X) \big[\phi^{\mu}_{\,\,\rho} \phi^{\rho \lambda} \phi_{\lambda}\\
&+&\phi^{\sigma} \phi_{\sigma \rho} \phi^{\rho \mu} \big]-\nabla_{\nu}\big[A_4(X) \big(\phi^{\mu} \phi^{\nu \rho} \phi_{\rho}\\
&+&\phi^{\sigma} \phi_{\,\,\,\sigma}^{\mu}
\phi^{\nu}\big)\big], \\ \\
\mathcal{J}^{\mu}_{(5)}&=&2 A_{5X} \phi^{\mu} \big( \phi^{\sigma} \phi_{\sigma
\rho} \phi^{\rho}\big)^2 +2 A_5(X)\big(
\phi^{\sigma} \phi_{\sigma \rho} \phi^{\rho}\big)\big(\phi^{\mu \sigma}\phi_{\sigma}\\
&+&\phi^{\sigma \mu} \phi_{\sigma}\big)-2\nabla_{\nu}\left[A_5(X)
\phi^{\sigma} \phi_{\sigma \rho} \phi^{\rho}  \phi^{\mu}
\phi^{\nu}\right].
\end{eqnarray*}


\begin{thebibliography}{}



\bibitem{Santos:2020xox}
  F.~F.~Santos,
  {\it Rotating black hole with a probe string in Horndeski Gravity},
  Eur.\ Phys.\ J.\ Plus {\bf 135}, no. 10, 810 (2020),
  [arXiv:2005.10983 [hep-th]].
	
\bibitem{Nagasaki:2017kqe}
K.~Nagasaki,
{\it Complexity of AdS$_5$ black holes with a rotating string},
Phys. Rev. D \textbf{96}, no.12, 126018 (2017),
[arXiv:1707.08376 [hep-th]].

\bibitem{Nagasaki:2018csh}
K.~Nagasaki,
{\it Complexity growth of rotating black holes with a probe string},
Phys. Rev. D \textbf{98}, no.12, 126014 (2018),
[arXiv:1807.01088 [hep-th]].

\bibitem{Roberts:2016hpo}
D.~A.~Roberts and B.~Yoshida,
{\it Chaos and complexity by design},
JHEP \textbf{04}, 121 (2017),
[arXiv:1610.04903 [quant-ph]].

\bibitem{Stanford:2014jda}
D.~Stanford and L.~Susskind,
{\it Complexity and Shock Wave Geometries},
Phys. Rev. D \textbf{90}, no.12, 126007 (2014),
[arXiv:1406.2678 [hep-th]].

\bibitem{Carmi:2017jqz}
D.~Carmi, S.~Chapman, H.~Marrochio, R.~C.~Myers and S.~Sugishita,
{\it On the Time Dependence of Holographic Complexity},
JHEP \textbf{11}, 188 (2017),
[arXiv:1709.10184 [hep-th]].

\bibitem{Susskind:2014rva}
L.~Susskind,
{\it Computational Complexity and Black Hole Horizons},
Fortsch. Phys. \textbf{64}, 24-43 (2016),
[arXiv:1403.5695 [hep-th]].

\bibitem{Watrous2008}
J. Watrous,
{\it Quantum Computational Complexity}, (2008),
[arXiv:quant-ph/0804.3401].

\bibitem{Brown:2017jil}
A.~R.~Brown and L.~Susskind,
{\it Second law of quantum complexity},
Phys. Rev. D \textbf{97}, no.8, 086015 (2018),
[arXiv:1701.01107 [hep-th]].

\bibitem{Almheiri:2012rt}
A.~Almheiri, D.~Marolf, J.~Polchinski and J.~Sully,
{\it Black Holes: Complementarity or Firewalls?},
JHEP \textbf{02}, 062 (2013),
[arXiv:1207.3123 [hep-th]].

\bibitem{Harlow:2013tf}
D.~Harlow and P.~Hayden,
{\it Quantum Computation vs. Firewalls},
JHEP \textbf{06}, 085 (2013),
[arXiv:1301.4504 [hep-th]].

\bibitem{Susskind:2015toa}
L.~Susskind,
{\it The Typical-State Paradox: Diagnosing Horizons with Complexity},
Fortsch. Phys. \textbf{64}, 84-91 (2016),
[arXiv:1507.02287 [hep-th]].

\bibitem{Brown:2015bva}
A.~R.~Brown, D.~A.~Roberts, L.~Susskind, B.~Swingle and Y.~Zhao,
{\it Holographic Complexity Equals Bulk Action?},
Phys. Rev. Lett. \textbf{116}, no.19, 191301 (2016),
[arXiv:1509.07876 [hep-th]].

\bibitem{Brown:2015lvg}
A.~R.~Brown, D.~A.~Roberts, L.~Susskind, B.~Swingle and Y.~Zhao,
{\it Complexity, action, and black holes},
Phys. Rev. D \textbf{93}, no.8, 086006 (2016),
[arXiv:1512.04993 [hep-th]].


\bibitem{Banados:1992wn}
M.~Banados, C.~Teitelboim and J.~Zanelli,
Phys. Rev. Lett. \textbf{69} (1992), 1849-1851
doi:10.1103/PhysRevLett.69.1849
[arXiv:hep-th/9204099 [hep-th]].


\bibitem{Horndeski:1974wa}
G.~W.~Horndeski,
{\it Second-order scalar-tensor field equations in a four-dimensional space},
Int.\ J.\ Theor.\ Phys.\  {\bf 10}, 363 (1974).



\bibitem{Maldacena:1997re}
J.~M.~Maldacena,
Adv. Theor. Math. Phys. \textbf{2} (1998), 231-252
doi:10.1023/A:1026654312961
[arXiv:hep-th/9711200 [hep-th]].





\bibitem{Gubser:1998bc}
S.~S.~Gubser, I.~R.~Klebanov and A.~M.~Polyakov,
Phys. Lett. B \textbf{428} (1998), 105-114
doi:10.1016/S0370-2693(98)00377-3
[arXiv:hep-th/9802109 [hep-th]].



\bibitem{Witten:1998qj}
E.~Witten,
Adv. Theor. Math. Phys. \textbf{2} (1998), 253-291
doi:10.4310/ATMP.1998.v2.n2.a2
[arXiv:hep-th/9802150 [hep-th]].




\bibitem{Iqbal:2008by}
N.~Iqbal and H.~Liu,
Phys. Rev. D \textbf{79} (2009), 025023
doi:10.1103/PhysRevD.79.025023
[arXiv:0809.3808 [hep-th]].

\bibitem{Brito:2019ose}
F.~A.~Brito and F.~F.~Santos,
{\it Black brane in asymptotically Lifshitz spacetime and viscosity/entropy ratios in Horndeski gravity},
EPL \textbf{129}, no.5, 50003 (2020),
[arXiv:1901.06770 [hep-th]].

\bibitem{Kovtun:2003wp}
  P.~Kovtun, D.~T.~Son and A.~O.~Starinets,
  {\it Holography and hydrodynamics: Diffusion on stretched horizons},
  JHEP {\bf 0310}, 064 (2003),
  [hep-th/0309213].

\bibitem{Kovtun:2004de}
P.~Kovtun, D.~T.~Son and A.~O.~Starinets,
{\it Viscosity in strongly interacting quantum field theories from black hole physics},
Phys. Rev. Lett. \textbf{94}, 111601 (2005),
[arXiv:hep-th/0405231 [hep-th]].


\bibitem{Son:2002sd}
D.~T.~Son and A.~O.~Starinets,
JHEP \textbf{09} (2002), 042
doi:10.1088/1126-6708/2002/09/042
[arXiv:hep-th/0205051 [hep-th]].

\bibitem{Buchel:2003tz}
A.~Buchel and J.~T.~Liu,
Phys. Rev. Lett. \textbf{93} (2004), 090602
doi:10.1103/PhysRevLett.93.090602
[arXiv:hep-th/0311175 [hep-th]].




\bibitem{Buchel:2004qq}
A.~Buchel,
Phys. Lett. B \textbf{609} (2005), 392-401
doi:10.1016/j.physletb.2005.01.052
[arXiv:hep-th/0408095 [hep-th]].



\bibitem{Benincasa:2006fu}
P.~Benincasa, A.~Buchel and R.~Naryshkin,
Phys. Lett. B \textbf{645} (2007), 309-313
doi:10.1016/j.physletb.2006.12.030
[arXiv:hep-th/0610145 [hep-th]].



\bibitem{Landsteiner:2007bd}
K.~Landsteiner and J.~Mas,
JHEP \textbf{07} (2007), 088
doi:10.1088/1126-6708/2007/07/088
[arXiv:0706.0411 [hep-th]].



\bibitem{Kats:2007mq}
Y.~Kats and P.~Petrov,
JHEP \textbf{01} (2009), 044
doi:10.1088/1126-6708/2009/01/044
[arXiv:0712.0743 [hep-th]].




\bibitem{Brigante:2007nu}
M.~Brigante, H.~Liu, R.~C.~Myers, S.~Shenker and S.~Yaida,
Phys. Rev. D \textbf{77} (2008), 126006
doi:10.1103/PhysRevD.77.126006
[arXiv:0712.0805 [hep-th]].






\bibitem{Feng:2015oea}
X.~H.~Feng, H.~S.~Liu, H.~L\"u and C.~N.~Pope,
JHEP \textbf{11} (2015), 176
doi:10.1007/JHEP11(2015)176
[arXiv:1509.07142 [hep-th]].

\bibitem{Fan:2018qnt}
Z.~Y.~Fan,
Phys. Rev. D \textbf{97} (2018) no.6, 066013
doi:10.1103/PhysRevD.97.066013
[arXiv:1801.07870 [hep-th]].


\bibitem{Motohashi:2016ftl}
H.~Motohashi, K.~Noui, T.~Suyama, M.~Yamaguchi and D.~Langlois,
JCAP \textbf{07} (2016), 033
doi:10.1088/1475-7516/2016/07/033
[arXiv:1603.09355 [hep-th]].

\bibitem{BenAchour:2016fzp}
J.~Ben Achour, M.~Crisostomi, K.~Koyama, D.~Langlois, K.~Noui and G.~Tasinato,
{\it Degenerate higher order scalar-tensor theories beyond Horndeski up to cubic order},
JHEP \textbf{12} (2016), 100,
[arXiv:1608.08135 [hep-th]].



\bibitem{Babichev:2020qpr}
E.~Babichev, C.~Charmousis, A.~Cisterna and M.~Hassaine,
{\it Regular black holes via the Kerr-Schild construction in DHOST theories},
JCAP \textbf{06} (2020), 049
doi:10.1088/1475-7516/2020/06/049
[arXiv:2004.00597 [hep-th]].

\bibitem{Baake:2021jzv}
O.~Baake, C.~Charmousis, M.~Hassaine and M.~San Juan,
JCAP \textbf{06} (2021), 021
doi:10.1088/1475-7516/2021/06/021
[arXiv:2104.08221 [hep-th]].


\bibitem{Baake:2021kyg}
O.~Baake and M.~Hassaine,
Eur. Phys. J. C \textbf{81} (2021), 642
doi:10.1140/epjc/s10052-021-09449-2
[arXiv:2104.13834 [hep-th]].



\bibitem{Baake:2020tgk}
O.~Baake, M.~F.~Bravo Gaete and M.~Hassaine,
{\it Spinning black holes for generalized scalar tensor theories in three dimensions},
Phys. Rev. D \textbf{102} (2020) no.2, 024088,
[arXiv:2005.10869 [hep-th]].


\bibitem{Kobayashi:2014eva}
T.~Kobayashi and N.~Tanahashi,
{\it Exact black hole solutions in shift symmetric scalar\textendash{}tensor theories},
PTEP \textbf{2014} (2014), 073E02,
[arXiv:1403.4364 [gr-qc]].





\bibitem{Babichev:2017guv}
E.~Babichev, C.~Charmousis and A.~Leh\'ebel,
JCAP \textbf{04} (2017), 027
doi:10.1088/1475-7516/2017/04/027
[arXiv:1702.01938 [gr-qc]].



\bibitem{Chagoya:2018lmv}
J.~Chagoya and G.~Tasinato,
JCAP \textbf{08} (2018), 006
doi:10.1088/1475-7516/2018/08/006
[arXiv:1803.07476 [gr-qc]].


\bibitem{Charmousis:2019vnf}
C.~Charmousis, M.~Crisostomi, R.~Gregory and N.~Stergioulas,
Phys. Rev. D \textbf{100} (2019) no.8, 084020
doi:10.1103/PhysRevD.100.084020
[arXiv:1903.05519 [hep-th]].


\bibitem{Babichev:2013cya}
E.~Babichev and C.~Charmousis,
{\it Dressing a black hole with a time-dependent Galileon},
JHEP \textbf{08} (2014), 106,
[arXiv:1312.3204 [gr-qc]].

\bibitem{Anabalon:2013oea}
A.~Anabalon, A.~Cisterna and J.~Oliva,
{\it Asymptotically locally AdS and flat black holes in Horndeski theory},
Phys.\ Rev.\ D {\bf 89}, 084050 (2014),
[arXiv:1312.3597 [gr-qc]].


\bibitem{Bravo-Gaete:2014haa}
M.~Bravo-Gaete and M.~Hassaine,
{\it Thermodynamics of a BTZ black hole solution with an Horndeski source},''
Phys. Rev. D \textbf{90} (2014) no.2, 024008,
[arXiv:1405.4935 [hep-th]].


\bibitem{Kim:2013zha}
  W.~Kim, S.~Kulkarni and S.~H.~Yi,
  {\it Quasilocal Conserved Charges in a Covariant Theory of Gravity},
  Phys.\ Rev.\ Lett.\  {\bf 111}, no. 8, 081101 (2013)
  [arXiv:1306.2138 [hep-th]].

\bibitem{Gim:2014nba}
  Y.~Gim, W.~Kim and S.~H.~Yi,
  {\it The first law of thermodynamics in Lifshitz black holes revisited},
  JHEP {\bf 1407}, 002 (2014)
  [arXiv:1403.4704 [hep-th]].

\bibitem{Abbott:1981ff}
  L.~F.~Abbott and S.~Deser,
  {\it Stability of Gravity with a Cosmological Constant},
  Nucl.\ Phys.\ B {\bf 195}, 76 (1982).

\bibitem{Deser:2002rt}
  S.~Deser and B.~Tekin,
  {\it Gravitational energy in quadratic curvature gravities},
  Phys.\ Rev.\ Lett.\  {\bf 89}, 101101 (2002)
  [hep-th/0205318].

\bibitem{Deser:2002jk}
  S.~Deser and B.~Tekin,
  {\it Energy in generic higher curvature gravity theories},
  Phys.\ Rev.\ D {\bf 67}, 084009 (2003),
  [hep-th/0212292].

\bibitem{Herrera-Aguilar:2020iti}
A.~Herrera-Aguilar, D.~F.~Higuita-Borja and J.~A.~M\'endez-Zavaleta,
{\it Black hole scalarization through spacetime anisotropic scaling symmetry},
[arXiv:2012.13412 [hep-th]].

\bibitem{Bravo-Gaete:2020ftn}
M.~Bravo-Gaete and M.~M.~Ju\'arez-Aubry,
{\it Thermodynamics and Cardy-like formula for nonminimally dressed, charged Lifshitz black holes in new massive gravity},
Class. Quant. Grav. \textbf{37} (2020) no.7, 075016,
[arXiv:2002.10520 [hep-th]].

\bibitem{Ayon-Beato:2019kmz}
E.~Ay\'on-Beato, M.~Bravo-Gaete, F.~Correa, M.~Hassaine and M.~M.~Ju\'arez-Aubry,
{\it Microscopic entropy of higher-dimensional nonminimally dressed Lifshitz black holes},''
Phys. Rev. D \textbf{100} (2019) no.4, 044024
doi:10.1103/PhysRevD.100.044024
[arXiv:1904.09391 [hep-th]].

\bibitem{BravoGaete:2017dso}
M.~Bravo Gaete, L.~Guajardo and M.~Hassaine,
JHEP \textbf{04} (2017), 092
doi:10.1007/JHEP04(2017)092
[arXiv:1702.02416 [hep-th]].

\bibitem{Bravo-Gaete:2021kgt}
M.~Bravo-Gaete, M.~M.~Juarez-Aubry and G.~Velazquez-Rodriguez,
[arXiv:2112.01483 [hep-th]].

\bibitem{Wald:1993nt}
  R.~M.~Wald,
  {\it Black hole entropy is the Noether charge},
  Phys.\ Rev.\ D {\bf 48}, no. 8, R3427 (1993),
  [gr-qc/9307038].

\bibitem{Iyer:1994ys}
  V.~Iyer and R.~M.~Wald,
  {\it Some properties of Noether charge and a proposal for dynamical black hole entropy},
  Phys.\ Rev.\ D {\bf 50}, 846 (1994),
  [gr-qc/9403028].
  
 
\bibitem{Smarr:1972kt}
L.~Smarr,
Phys. Rev. Lett. \textbf{30} (1973), 71-73
[erratum: Phys. Rev. Lett. \textbf{30} (1973), 521-521]
doi:10.1103/PhysRevLett.30.71
	
\bibitem{Abad:2017cgl}
F.~J.~G.~Abad, M.~Kulaxizi and A.~Parnachev,
{\it On Complexity of Holographic Flavors},
JHEP \textbf{01}, 127 (2018),
[arXiv:1705.08424 [hep-th]].

\bibitem{Bravo-Gaete:2021hlc}
M.~Bravo-Gaete and M.~M.~Stetsko,
{\it Planar black holes configurations and shear viscosity in arbitrary dimensions with shift and reflection symmetric scalar-tensor theories},
[arXiv:2111.10925 [hep-th]].

\bibitem{Lehner:2016vdi}
L.~Lehner, R.~C.~Myers, E.~Poisson and R.~D.~Sorkin,
{\it Gravitational action with null boundaries},
Phys. Rev. D \textbf{94} (2016) no.8, 084046,
[arXiv:1609.00207 [hep-th]].





\end{thebibliography}
\end{document}